\documentclass[12pt]{article}
\usepackage{amsmath} 
\usepackage{amsthm}
\usepackage{amssymb}
\usepackage{amsopn}

\newcommand{\ssection}[1]{ 
\refstepcounter{section}
\setcounter{equation}{0} 
\setcounter{subsection}{0} 
\addcontentsline{toc}{section} 
      {\normalsize\textbf{\thesection.\ #1}} 
\bigskip\bigskip\noindent 
\normalsize\textbf{\thesection.\ #1}\nopagebreak\smallskip\nopagebreak} 
\def\thesection{{\normalsize\arabic{section}}} 
\newcommand{\ssubsection}[1]{ 
\refstepcounter{subsection}
\addcontentsline{toc}{subsection} 
      {\normalsize\normalfont\textit{\thesubsection.\ #1}} 
\medskip\medskip\noindent 
\normalsize\normalfont
\textit{\thesubsection. \ #1}\nopagebreak\smallskip\nopagebreak} 
\def\thesubsection{{\normalsize
{\arabic{section}.\arabic{subsection}}}} 
\newcounter{appendice}
\def\theappendice{{\normalsize\Alph{appendice}}} 
\newcommand{\appendice}[1]{
\refstepcounter{appendice}
\setcounter{equation}{0} 
\def\theequation{\theappendice.\arabic{equation}} 
\addcontentsline{toc}{section} 
      {\normalsize\textbf{\theappendice.\ #1}} 
\bigskip\bigskip\noindent 
\normalsize\textbf{\theappendice.\ #1}\par\smallskip\nopagebreak} 

\def\theequation{\arabic{section}.\arabic{equation}}

\newcommand{\id}{{1 \mskip -5mu {\rm I}}}
\newcommand{\reff}[1]{(\ref{#1})}

\newcommand{\mf}[1]{{\mathfrak #1}}

\newcommand{\bb}[1]{{\mathbb #1}}

\begin{document}

\begin{titlepage}

\par\vskip 1cm\vskip 2em

\begin{center}

{\LARGE  Macroscopic fluctuation theory \\ ~\\ 
for stationary non  equilibrium states }

\par

\vskip 2.5em \lineskip .5em

{\large
\begin{tabular}[t]{c}
$\mbox{L. Bertini}^{1}\!\!\phantom{m}\mbox{ A. De Sole}^{2}
\!\phantom{m}\mbox{D. Gabrielli}^{3}\!\phantom{m}
\mbox{G. Jona--Lasinio}^{4}\!\phantom{m}\mbox{C. Landim}^{5}$ 
\\
\end{tabular}
\par
}
\medskip
{\small
\begin{tabular}[t]{ll}
{\bf 1} & 
Dipartimento di Matematica, Universit\`a di Roma La Sapienza\\
&  P.le A.\ Moro 2, 00185 Roma, Italy\\
&  E--mail: {\tt lorenzo@carpenter.mat.uniroma1.it}\\
{\bf 2} & Department of Mathematics, MIT \\
& 77 Massachusetts Avenue, Cambridge, MA 02139-4307, USA \\
& E--mail: {\tt desole@math.mit.edu}\\
{\bf 3} & Dipartimento di Matematica, Universit\`a dell'Aquila\\
&  67100 Coppito, L'Aquila, Italy \\
&  E--mail: {\tt gabriell@univaq.it}\\
{\bf 4} & Dipartimento di Fisica and INFN, Universit\`a di Roma La Sapienza\\
&  P.le A.\ Moro 2, 00185 Roma, Italy\\
&  E--mail: {\tt jona@roma1.infn.it}\\ 
{\bf 5}& IMPA, Estrada Dona Castorina 110, J. Botanico, 22460 Rio
de Janeiro, Brazil\\
& CNRS UPRES--A 6085, Universit\'e de Rouen,
76128 Mont--Saint--Aignan Cedex, France \\
& E--mail: {\tt landim@impa.br}\\
\end{tabular}
}
\bigskip
\end{center}

\vskip 1 em
\centerline{\bf Abstract} 
\smallskip
{\small 
\noindent
We formulate a dynamical fluctuation theory for stationary non
equilibrium states (SNS) which is tested explicitly in stochastic
models of interacting particles.  In our theory a crucial role is
played by the time reversed dynamics. Within this theory we derive the
following results: the modification of the Onsager--Machlup theory in
the SNS; a general Hamilton--Jacobi equation for the macroscopic
entropy; a non equilibrium, non linear fluctuation dissipation
relation valid for a wide class of systems; an H theorem for the
entropy.  We discuss in detail two models of stochastic boundary
driven lattice gases: the zero range and the simple exclusion
processes.  In the first model the invariant measure is explicitly
known and we verify the predictions of the general theory. For the one
dimensional simple exclusion process, as recently shown by Derrida,
Lebowitz, and Speer, it is possible to express the macroscopic entropy
in terms of the solution of a non linear ordinary differential
equation; by using the Hamilton--Jacobi equation, we obtain a logically 
independent derivation of this result.
}

\vfill


\end{titlepage}
\vfill\eject

\ssection{Introduction}
\par\noindent
The Boltzmann--Einstein theory of equilibrium thermodynamic
fluctuations, as described for example in Landau--Lifshitz \cite{LL},
states that the probability for a fluctuation from equilibrium in a
macroscopic region of volume $V$ is proportional to 
$$ 
\exp\{V\Delta S / k\}
$$ 
where $\Delta S$ is the variation of entropy density calculated along
a reversible transformation creating the fluctuation and $k$ is the
Boltzmann constant. This theory is well established and has received a
rigorous mathematical formulation in classical equilibrium statistical
mechanics via the so called large deviation theory \cite{La}.  The
rigorous study of large deviations has been extended to hydrodynamic
evolutions of stochastic interacting particle systems \cite{kov}. In a
dynamical setting one may ask new questions, for example what is the
most probable trajectory followed by the system in the spontaneous
emergence of a fluctuation or in its relaxation to equilibrium.  The
Onsager--Machlup theory \cite{ON2} gives the following answer under
the assumption of time reversibility.  In the situation of a linear
macroscopic equation, that is, close to equilibrium, the most probable
emergence and relaxation trajectories are one the time reversal of the
other. Developing the methods of \cite{kov}, this theory has been
extended to nonlinear hydrodynamic regimes \cite{JLV}.

In the present paper we formulate a general theory of large deviations
for irreversible processes, i.e.\ when detailed balance does not
hold. This question was previously addressed in \cite{GT} for finite
dimensional diffusions and in \cite{EY} for lattice gases.  Natural
examples are boundary driven stationary non equilibrium states (SNS),
e.g.\ a thermodynamic system in contact with two reservoirs.  In such
a situation there is a flow of matter or other physical property like
heat, charge,...  through the system.  As we shall see, the
spontaneous fluctuations of the process are described by the time
reversed dynamics, which is defined below.

Spontaneous fluctuations, including Onsager--Machlup symmetry, have
been observed in stochastically perturbed reversible electronic
devices \cite{LM}.  In their work, these authors study also non
reversible systems and observe violation of Onsager--Machlup symmetry;
in the present work we shall connect such violations to the time
reversed dynamics.

\medskip
We are interested in many body systems in the limit of infinitely many
degrees of freedom. The basic assumptions of our theory are the
following.

1) The microscopic evolution is given by a Markov process $X_t$ which
represents the configuration of the system at time $t$. 
This hypothesis probably is not so restrictive because also the
Hamiltonian case discussed in \cite{EPR} in the end is reduced to the
analysis of a Markov process. The stationary non equilibrium state
(SNS) is described by a stationary, i.e.\ invariant with respect to
time shifts, probability distribution $P_{st}$ over the trajectories of
$X_t$.

2) The system admits also a macroscopic description in terms of
density fields which are the local thermodynamic variables.  For
simplicity of notation we assume there is only one
thermodynamic variable $\rho$.  The evolution of the field
$\rho=\rho(t,u)$ where $u$ is the macroscopic space coordinate, is
given by a diffusion type hydrodynamical equation of the form
\begin{equation}
\partial_t \rho  
=\frac 12 \nabla \cdot \big( D(\rho)\nabla \rho \big)
=\frac 12 \sum_{1\le i,j \le d}\partial_{u_i}
\big( D_{i,j}(\rho)\, \partial_{u_j} \rho \big)
= \mf D (\rho)
\label{H}
\end{equation}
The interaction with the reservoirs appears as boundary conditions to
be imposed on solutions of (\ref{H}).  We assume that there exists a
unique stationary solution $\bar\rho$ of (\ref{H}), i.e.\ a profile
$\bar\rho(u)$, which satisfies the appropriate boundary conditions
such that $\mf D(\bar\rho)=0$. This holds if the diffusion matrix
$D_{i,j}(\rho)$ in (\ref{H}) is strictly elliptic, namely there exists
a constant $c>0$ such that $D(\rho)\ge c\id $ (in matrix
sense).

This equation derives from the underlying microscopic dynamics
through an appropriate scaling limit.  The hydrodynamic equation
\reff{H} represents a law of large numbers with respect to the
probability measure $P_{st}$ conditioned on an initial state
$X_0$. The initial condition for (\ref{H}) is determined by
$X_0$. Of course many microscopic configurations give rise to the same
value of $\rho(0,u)$. In general $\rho=\rho(t,u)$ is an appropriate
limit of a $\rho_N(X_t)$ as the number $N$ of degrees of freedom
diverges.

3) Let us denote by $\theta$ the time inversion operator defined by
$\theta X_t = X_{-t} $. The probability measure $P^*_{st}$
describing the evolution of the time reversed process $ X_{t}^*$ is
given by the composition of $P_{st}$ and ${\theta}^{-1}$ that is
\begin{equation}
\label{P*}
P^*_{st} (X_t^*=\phi_t, t\in [t_1,t_2]) = 
P_{st} (X_t=\phi_{-t}, t\in [-t_2,-t_1])
\end{equation}

Let $L$ be the generator of the microscopic dynamics.  We remind that
$L$ induces the evolution of observables (functions on the
configuration space) according to the equation $\partial_t
E_{X_0}[f(X_t)] = E_{X_0}[(Lf)(X_t)]$, where $E_{X_0}$ stands for the
expectation with respect to $P_{st}$ conditioned on the initial state
$X_0$, see e.g.\ \cite[Ch. X]{FE}.  The time reversed dynamics is
generated by the adjoint $L^*$ of $L$ with respect to the invariant
measure $\mu$, that is
\begin{equation}
\label{L*}
E^{\mu}[fLg]= E^{\mu}[(L^*f)g]
\end{equation}
The measure $\mu$, which is the same for both processes, is a
distribution over the configurations of the system and formally
satisfies $\mu L=0$.  The expectation with respect to $\mu$ is denoted
by $E^{\mu}$ and $f,g$ are observables.  We note that the probability
$P_{st}$, and therefore $P^*_{st}$, depends on the invariant measure
$\mu$. The finite dimensional distributions of $P_{st}$ are in fact
given by
\begin{equation} 
P_{st}\left( X_{t_1} =\phi_{t_1}, \cdots, X_{t_n}=\phi_{t_n}\right)
= \mu ( \phi_{t_1}) \, p_{t_2-t_1}( \phi_{t_1} \rightarrow  \phi_{t_2})
\, \cdots  \, p_{t_n-t_{n-1}}( \phi_{t_{n-1}} \rightarrow  \phi_{t_n}) 
\end{equation}
where $p_t(\phi_1\rightarrow \phi_2)$ is the transition
probability. According to \reff{P*} the finite dimensional
distributions of $P_{st}^*$ are
\begin{eqnarray} 
&&\!\!\!\!\!\!
P_{st}^*\left( X_{t_1}^* =\phi_{t_1}, \cdots, X_{t_n}^*=\phi_{t_n}\right)
= \mu ( \phi_{t_1}) \, 
p^*_{t_2-t_1}( \phi_{t_1} \rightarrow  \phi_{t_2})
\, \cdots \, p^*_{t_n-t_{n-1}}( \phi_{t_{n-1}} \rightarrow  \phi_{t_n}) 
\phantom{mer}
\nonumber\\
&&\!\!\!\!\!\!\phantom{P_{st}^* (........)} = \mu(\phi_{t_n}) 
p_{t_n-t_{n-1}} ( \phi_{t_n}\rightarrow \phi_{t_{n-1}} )
\, \cdots  \, p_{t_2-t_{1}} ( \phi_{t_2}\rightarrow \phi_{t_1} )
\end{eqnarray}
in particular the transition probabilities 
$p_t(\phi_1\rightarrow \phi_2)$ and 
$p_t^*(\phi_1\rightarrow \phi_2)$ are related by
\begin{equation}
\mu(\phi_1)\, p_t(\phi_1\rightarrow \phi_2) 
= \mu(\phi_2) \, p_t^*(\phi_2 \rightarrow \phi_1 )
\end{equation}
which reduces to the well known detailed balance condition if 
$p_t(\phi_1\rightarrow \phi_2)=p_t^*(\phi_1\rightarrow \phi_2)$.

We require that also the evolution generated by $L^*$ admits a
hydrodynamic description, that we call the adjoint hydrodynamics,
which, however, is not necessarily of the same form as (\ref{H}).  In
fact we shall discuss a model in which the adjoint hydrodynamics is
non local in space.

In order to avoid confusion we emphasize that what is usually called 
an equilibrium state, as distinguished from a SNS, corresponds to the
special case $L^* = L$, i.e.\ the detailed balance principle holds.  
In such a case $P_{st}$ is invariant under time reversal and 
the two hydrodynamics coincide.

4) The stationary measure $P_{st}$ admits a principle of large deviations
describing the fluctuations of the thermodynamic variable appearing
in the hydrodynamic equation. This means the following. 
The probability that in a macroscopic volume $V$ containing
$N$ particles  the evolution of the variable $\rho_N$
deviates from the solution of the hydrodynamic equation and is close
to some trajectory ${\hat{\rho}}(t)$, is exponentially
small and of the form
\begin{equation}
\label{LD}
P_{st}\left( 
\rho_N(X_t) \sim\hat{\rho}(t), t\in [t_1, t_2]\right) 
\approx e^{-N[S(\hat \rho({t_1})) + J_{[t_1,t_2]}(\hat{\rho})] } 
= e^{-N I_{[t_1,t_2]}(\hat{\rho})}
\end{equation}
where $J(\hat{\rho})$ is a functional which vanishes if
${\hat{\rho}}(t)$ is a solution of (\ref{H}) and $S(\hat \rho({t_1}))$
is the entropy cost to produce the initial density profile
${\hat{\rho}}({t_1})$.  We adopt the convention for the entropy sign
opposite to the usual one, so that it takes the minimum value in the
equilibrium state. We also normalize it so that $S(\bar\rho)=0$.
The functional $J(\hat{\rho})$ represents the extra cost necessary to
follow the trajectory ${\hat{\rho}}(t)$. Finally $\rho_N(X_t) \sim
\hat{\rho}(t)$ means closeness in some metric and $\approx$ denotes
logarithmic equivalence as $N\to\infty$.  This formula is a
generalization of the Boltzmann--Einstein. We set the
Boltzmann constant $k=1$.

\bigskip
This paper is divided in two parts. In the first one
we present a general fluctuation theory of SNS based on the hypotheses
formulated above assuming the knowledge of the functional
$J(\hat{\rho})$. The main results are outlined below.

\begin{enumerate}

\item{The Onsager--Machlup relationship has to be modified in the
following way: {\sl{``In a SNS the spontaneous emergence of a
macroscopic fluctuation takes place most likely following a trajectory
which is the time reversal of the relaxation path according to the
adjoint hydrodynamics"}}.}

\item{
We show that the macroscopic entropy $S(\rho)$ solves a
Hamilton--Jacobi equation generalizing to a thermodynamic context
known results for finite dimensional Lan\-ge\-vin equations 
\cite{GT,FW,LM}.}

\item{
For a large class of systems we obtain a non equilibrium non
linear fluctuation dissipation relationship which links the
hydrodynamic evolutions of the system and of its time reversal
to the thermodynamic force, that is the derivative of $S(\rho)$. 
If $S(\rho)$ is known this relationship determines the adjoint
hydrodynamics. 
}

\item{From the last two results we derive an H Theorem for $S(\rho)$:
it is decreasing along the solutions of both the hydrodynamics and the
adjoint  hydrodynamics.
}

\end{enumerate}

\medskip
In the second part we test the theory outlined above in 
two boundary driven stochastic models of interacting particle
systems: the zero range and the simple exclusion processes. 
The main results are outlined below.

\begin{enumerate}
\item{ For the boundary driven zero range process, as observed in
\cite{DF}, the invariant measure is a product measure. It is therefore
possible to write the functional $S(\rho)$, which in this case is a
local functional of $\rho$, in a closed form and to construct the
microscopic time reversed process explicitly.  We derive both the
hydrodynamics and the adjoint hydrodynamics. We compute the
functionals $J(\hat\rho)$ and $J^*(\hat\rho)$ and verify the
generalized Onsager--Machlup principle and the fluctuation dissipation
relationship.  }

\item{
For the boundary driven simple exclusion process the invariant
measure has long range correlations and is not explicitly known.
The hydrodynamics has been obtained in \cite{els1,els2}; we obtain the
asymptotics of the probability of large deviations, that is we calculate
$J(\hat\rho)$.  
In one space dimension, Derrida, Lebowitz, and Speer \cite{DLS} have
recently shown that the action functional $S$ can be expressed in
terms of the solution of a non--linear ordinary differential
equation. We show how this result can be recovered by our approach: 
the Hamilton--Jacobi equation for $S(\rho)$ (which is a
functional derivative equation) can be reduced to the solution of the
ordinary differential equation obtained in \cite{DLS}.
By using the fluctuation dissipation relationship we also
find the adjoint hydrodynamics.  Moreover, in any spatial dimension,
we can deduce a lower bound for the macroscopic entropy
in terms of the entropy of an equilibrium state.  In the
one dimensional case this bound has been independently obtained in
\cite{DLS}.  }
\end{enumerate}

Part of the results presented here have been briefly reported in
\cite{NOI}. Rigorous mathematical treatment of the boundary driven
simple exclusion process will be given in \cite{NOImat}.

\bigskip
We conclude with some remarks to clarify the differences between
equilibrium and non equilibrium states.  The main problem in the SNS
derives from the following situation.  In equilibrium states the
thermodynamic properties are determined by the Gibbs distribution
which is specified by the Hamiltonian without solving a dynamical
problem. On the contrary, in a SNS we cannot construct the appropriate
ensemble without calculating first the invariant measure. At
thermodynamic level, we do not need all the information carried by the
invariant measure, but only the functional $S(\rho)$ appearing in the
generalized Boltzmann--Einstein formula (\ref{LD}).  In general
$S(\rho)$, contrary to the equilibrium case, is a non local functional
of the profile $\rho$.  It turns out that the entropy $S(\rho)$ can be
obtained, both in equilibrium and non equilibrium, from
$J(\hat{\rho})$, which is therefore the basic object of the
macroscopic theory. This step is simple for equilibrium, but highly
non trivial in non equilibrium.

\bigskip\bigskip\bigskip\bigskip
\ssection{General theory}
\label{s:gt}

\ssubsection{Generalized Onsager--Machlup relationship} 
\par\noindent
We now derive a first consequence of our assumptions, that is
the relationship between the action functionals $I$ and $I^*$
associated to the dynamics $L$ and $L^*$.  From equation (\ref{P*})
and our assumptions it follows that $P^*_{st}$ also admits a large
deviation principle with functional $I^*$ given by
\begin{equation}
I^*_{[t_1,t_2]}(\hat \rho) = I_{[-t_2,-t_1]}(\theta \hat \rho)
\label{I} 
\end{equation}
with obvious notations. More explicitly this equation reads
\begin{equation}
S({\hat \rho}({t_1})) + J^*_{[t_1,t_2]}(\hat \rho) =
S({\hat \rho}({t_2})) + J_{[-t_2,-t_1]}(\theta \hat \rho) 
\label{J}
\end{equation}
where ${\hat \rho}({t_1}), {\hat \rho}({t_2})$ are the initial and final
points of the trajectory and $S({\hat \rho}({t_i}))$ the entropies
associated with the creation of the fluctuations $\hat {\rho}({t_i})$
starting from the SNS. 
The functional $J^*$ vanishes on the solutions of the adjoint
hydrodynamics.  From (\ref{J}) we can  obtain the
generalization of Onsager-Machlup relationship for SNS.

The physical situation we are considering is the following.  The
system is macroscopically in the stationary state $\bar\rho$ at
$t=-\infty$ but at $t=0$ we find it in the state $\rho$.  We
want to determine the most probable trajectory followed in the
spontaneous creation of this fluctuation. According to (\ref{LD}) this
trajectory is the one that minimizes $J$ among all trajectories
$\hat\rho(t)$ connecting $\bar\rho$ to ${\rho}$ in the time interval
$[-\infty,0]$. From (\ref{J}), recalling that $S(\bar\rho)=0$, we have
that
\begin{equation}
J_{[-\infty,0]}(\hat \rho) = S(\rho) 
+ J^*_{[0,\infty]}(\theta \hat \rho) 
\label{OM}
\end{equation}
The right hand side is minimal if $J^*_{[0,\infty]}(\theta \hat
\rho)=0$ that is if $\theta \hat \rho$ is a solution of the adjoint
hydrodynamics.  The existence of such a relaxation solution is due
to the fact that the stationary solution $\bar\rho$ is attractive also
for the adjoint hydrodynamics.  We have therefore the following
generalization of Onsager--Machlup to SNS

{\sl{``In a SNS the spontaneous emergence of a macroscopic fluctuation
takes place most likely following a trajectory which is the time
reversal of the relaxation path according to the adjoint hydrodynamics"}}

We note that the reversibility of the microscopic process $X_t$, which
we call microscopic reversibility, is not needed in order to deduce
the classical Onsager--Machlup principle (i.e.\ that the trajectory which
creates the fluctuation is the time reversal of the relaxation
trajectory). In fact the classical Onsager--Machlup principle holds if
and only if the hydrodynamics coincides with the adjoint hydrodynamics,
which we call macroscopic reversibility. Indeed, it is possible to
construct microscopic non reversible models in which the classical 
Onsager--Machlup principle holds, see \cite{GJL1,GJL2,GJLV}.

\ssubsection{The Hamilton--Jacobi equation for the entropy}
\par\noindent
We assume that the functional $J$ has a density (which
plays the role of a Lagrangian), i.e.\
\begin{equation}
\label{Jj}
J_{[t_1,t_2]} (\hat\rho) \;=\; \int_{t_1}^{t_2} dt\,
\mathcal{L} \left( \hat\rho(t),\partial_t \hat\rho(t) \right) 
\end{equation} From (\ref{OM}) we have that the entropy is related to $J$ by  
\begin{equation}
\label{l1}
S(\rho)= \inf_{\hat \rho}  J_{[-\infty,0]}(\hat \rho)
\end{equation}
where the minimum is taken over all trajectories $\hat \rho (t)$ 
connecting $\bar\rho$ to $\rho$. Therefore $S$ must satisfy the
Hamilton--Jacobi equation associated to the action functional $J$.
Let us introduce the Hamiltonian $\mathcal{H}(\rho,H)$ as the
Legendre transform of $\mathcal{L}(\rho,\partial_t \rho)$, i.e.\
\begin{equation}
\label{lHJ}
\mathcal{H}(\rho, H) = \sup_{\xi} \left\{ \langle \xi, H \rangle  
- \mathcal{L}( \rho, \xi ) \right\}
\end{equation}
where $\langle \cdot,\cdot \rangle$ denotes integration with respect
to the macroscopic space coordinates $u$, that is the inner product in
$L_2(du)$. This notation will be employed throughout the whole paper.

Noting that $\mathcal{H}(\bar\rho, 0)=0$, the
Hamilton--Jacobi equation associated to (\ref{l1}) is
\begin{equation}
\label{l2}
\mathcal{H}\Big(\rho, \frac{\delta S}{\delta \rho} \Big) = 0
\end{equation}

This is an equation for the functional derivative $A(\rho)={\delta
S}/{\delta \rho}$ but not all the solutions of the equation
$\mathcal{H}\left(\rho,A(\rho)\right)=0$ are the derivative of some
functional. Of course only those which are the derivative of some
functional are relevant for us.  Furthermore, as well known in
mechanics, the Hamilton--Jacobi equation (\ref{l2}) has many
solutions and we shall discuss later the criterion to select the
correct one.

Let also introduce the pressure $G=G(h)$, where $h=h(u)$ can be
interpreted as a chemical potential profile, as the Legendre transform
of the entropy $S(\rho)$, namely
\begin{equation}\label{pressure}
G(h) = \sup_{\rho} \left\{ \langle h, \rho \rangle - S( \rho) \right\}
\end{equation}
Then, by Legendre duality, we have $\delta G/\delta h = \rho$ and 
$\delta S /\delta \rho = h $ so that $G(h)$ satisfies the dual
Hamilton--Jacobi equation 
\begin{equation}
\label{dualHJ}
\mathcal{H}\Big(\frac{\delta G}{\delta h}, h \Big) = 0
\end{equation}

\smallskip
We now specify the Hamilton-Jacobi equation (\ref{l2}) for boundary
driven lattice gases.  We assume that the large deviation functional
$J$ may be expressed as a quadratic functional of $\partial_t
\hat\rho$
\begin{eqnarray}
\label{J2}
&&J_{[t_1,t_2]} (\hat\rho) =
\frac 12 \int_{t_1}^{t_2} dt\,
\left\langle 
\nabla^{-1} \Big( \partial_t \hat\rho -\mf D (\rho)\Big) , 
\chi(\hat\rho)^{-1} \,
\nabla^{-1} \Big( \partial_t \hat\rho - \mf D (\rho)\Big)
\right\rangle 
\\
&&\;\;=\frac 12 \int_{t_1}^{t_2} dt\,
\left\langle \nabla^{-1} \Big(
\partial_t \hat\rho - \frac 12 \nabla \cdot 
\left( D(\hat\rho) \nabla \hat\rho \right)
\Big) , \chi(\hat\rho)^{-1} \,
\nabla^{-1} \Big(
\partial_t \hat\rho - \frac 12 \nabla \cdot 
\left( D(\hat\rho) \nabla \hat\rho \right)
\Big)\right\rangle
\nonumber
\end{eqnarray}
where $D(\rho) = D_{i,j} (\rho)$ is the diffusion matrix in the
hydrodynamic equation (\ref{H}) and by $\nabla^{-1} f$ we mean a
vector field whose divergence equals $f$.
The form \reff{J2}, which we derive in the models discussed below, is
expected to be very general; the functional $J(\hat\rho)$ measures how
much $\hat\rho$ differs from a solution of the hydrodynamics (\ref{H})
and the matrix $\chi(\rho) = \chi_{i,j}(\rho)$ reflects the intensity
of the fluctuations. See \cite[II. 3.7]{Slib} for a heuristic
derivation of (\ref{J2}) for reversible lattice gases.  This form of
$J$ is also typical for diffusion processes described by finite
dimensional Langevin equations \cite{FW}.

In this case the Lagrangian $\mathcal{L}$ is quadratic in
$\partial_t\hat\rho(t)$ and the associated Hamiltonian is given by
\begin{equation}
\label{Ham}
\mathcal{H}(\rho,H) = \frac 12 
\left\langle \nabla H, \chi(\rho) \nabla H \right\rangle
+ \frac 12 \left\langle H ,\nabla \cdot 
\big(  D(\rho) \nabla\rho \big) \right\rangle
\end{equation}
so that the Hamilton--Jacobi equation (\ref{l2}) takes the form
\begin{equation}
\label{HJ}
\Big \langle \nabla \frac {\delta S}{\delta \rho}  , 
\chi (\rho) \nabla \frac {\delta S}{\delta \rho} \Big \rangle
+ \Big \langle \frac {\delta S}{\delta \rho} , 
\nabla \cdot \big( D(\rho) \nabla \rho \big) \Big\rangle = 0
\end{equation}

We remark that the macroscopic entropy $S$, given by the
variational principle \reff{l1}, depends only on the action functional
$J$ and is therefore stable with respect to microscopic perturbations
which do not affect the dynamical large deviations.

\ssubsection{The adjoint hydrodynamics and a non equilibrium
fluctuation dissipation relation} 
\par\noindent
By assuming the quadratic form (\ref{J2}) also for $J^*$, we deduce a
twofold generalization of the celebrated fluctuation dissipation
relationship: it is valid in non equilibrium states and in non linear
regimes.

Such a relationship will hold provided the rate function 
$J^*$ of the time reversed process is of the form 
(\ref{J2}) with a different hydrodynamic equation (the adjoint
hydrodynamics) that we write in general as 
\begin{equation}
\partial_t \rho^* = \mf D^* (\rho^*)
\label{D*strano}
\end{equation}
with the same boundary conditions as (\ref{H}). 

We then assume $J^*$ has the form
\begin{equation}
J_{[t_1,t_2]}^*(\hat\rho)={\frac {1}{2}}\int_{t_1}^{t_2} \!dt\: 
 \left\langle( 
\nabla^{-1}\left(  \partial_t{\hat\rho}- \mf D^*(\hat\rho) \right)\,,\,
\chi(\hat\rho)^{-1} 
\nabla^{-1}\left(  \partial_t{\hat\rho}- \mf D^*(\hat\rho) \right)
\right\rangle
\label{J2*}
\end{equation}
By taking the variation of the equation (\ref{J}), we get
\begin{equation}
\label{gfd}
\mf D(\rho) +\mf D^*(\rho) =  
\nabla \cdot \Big( \chi(\rho) \nabla \frac {\delta S}{\delta \rho}
\Big)
\end{equation}
This relation holds for the non--equilibrium zero range process which
we discuss later.  We also note that it holds for the equilibrium
reversible models for which the large deviation principle has been
rigorously proven such as the simple exclusion process \cite{kov}, the
Landau--Ginzburg model \cite{DV} and its non--gradient version
\cite{Q}.  It is also easy to check that the linearization of
(\ref{gfd}) around the stationary profile $\bar \rho$ yields a
fluctuation dissipation relationship which reduces to the usual one in
equilibrium. Accordingly, the matrix $\chi(\rho)$ coincides with the
Onsager matrix as defined in \cite{GJL1,GJL2,Slib}.

In order to verify the fluctuation dissipation relation \reff{gfd},
we need $\mf D(\rho)$, $\mf D^*(\rho)$ and $ {\delta S} / {\delta
\rho}$. On the other hand, it can be used to obtain the adjoint
hydrodynamics from $\mf D(\rho)$ and $ {\delta S} / {\delta \rho}$; the
first is typically known and the second can be calculated from the
Hamilton--Jacobi equation \reff{HJ}.

\smallskip
Suppose we can decompose the hydrodynamics as the sum of a gradient of
some functional $V$ and a vector field $\mathcal{A}$ orthogonal to it in
the metric induced by the operator $K^{-1}$ where $K f = - \nabla
\cdot \big( \chi(\rho) \nabla f \big)$, namely
$$
\mf D(\rho) =  \frac 12 
\nabla \cdot 
\Big( \chi(\rho) \nabla \frac {\delta V}{\delta \rho} \Big) 
+ \mathcal{A}(\rho)
$$
with
$$
\Big\langle K\frac{\delta V}{\delta \rho} \, , K^{-1}\,
\mathcal{A}(\rho) \Big\rangle = \Big\langle \frac{\delta V}{\delta
\rho} \, ,\, \mathcal{A}(\rho) \Big\rangle = 0
$$
If ${\delta V}/{\delta \rho}$, like the thermodynamic force 
${\delta S}/{\delta \rho}$, vanishes at the boundary, it is easy to
check that the functional $V$ solves the Hamilton--Jacobi equation.

Conversely, given $S(\rho)$, by using the fluctuation dissipation
relationship \reff{gfd}, we can introduce a vector field
$\mathcal{A}(\rho)$ such that
\begin{eqnarray*}
\mf D(\rho) &=&  \frac 12 
\nabla \cdot 
\Big( \chi(\rho) \nabla \frac {\delta S}{\delta \rho} \Big) 
+ \mathcal{A}(\rho)
\\
\mf D^*(\rho) &=&  \frac 12 
\nabla \cdot 
\Big( \chi(\rho) \nabla \frac {\delta S}{\delta \rho} \Big) 
- \mathcal{A}(\rho)
\end{eqnarray*}
and the Hamilton--Jacobi equation implies the orthogonality condition 
$$
\Big\langle  \frac {\delta S}{\delta \rho} \,,\, \mathcal{A}(\rho)
\Big\rangle = 0
$$
Note the analogy with \cite[Thm IV.3.1]{FW} for diffusion processes.

\ssubsection{H Theorem} 
\par\noindent
We show that the functional $S$ is decreasing along the solutions of
both the hydrodynamic equation \reff{H} and the adjoint hydrodynamics
\begin{equation}
\label{adjH}
\partial_t \rho = \mf D^* (\rho) = 
\nabla\cdot \Big( \chi (\rho) \nabla
\frac{\delta S}{\delta \rho} \Big) - \mf D(\rho)\;
\end{equation}

Let $\rho(t)$ be a solution of \reff{H} or \reff{adjH}, 
by using the Hamilton--Jacobi equation \reff{HJ}, we get
\begin{equation}
\label{HT}
\frac{d}{dt} S(\rho(t)) = \Big\langle 
\frac{\delta S}{\delta \rho} (\rho(t)), \partial_t \rho(t) \Big\rangle
= -\frac{1}{2} 
\Big \langle \nabla \frac {\delta S}{\delta \rho} (\rho(t)) \, ,\, 
\chi (\rho(t)) \nabla \frac {\delta S}{\delta \rho}  (\rho(t))\Big \rangle
\le 0
\end{equation}
In particular we have that $\frac{d}{dt} S(\rho(t))=0$ if and only if
$\frac {\delta S}{\delta \rho}(\rho(t))=0$. Since we assumed there exists a
unique stationary profile $\bar\rho$, this implies that $\bar\rho$
is globally attractive also for the adjoint hydrodynamics \reff{adjH}.

\ssubsection{A lower bound for the entropy $S$}
\par\noindent
Let us consider any functional $V(\rho)$, normalized so that
$V(\bar\rho)=0$, whose derivative satisfies the Hamilton--Jacobi
equation (\ref{HJ}) and the condition $\delta V(\rho)/\delta \rho =0$
at the boundary.  We shall prove the bound
\begin{equation}
\label{lS>V}
S(\rho) =  \inf_{\hat \rho}  J_{[-\infty,0]}(\hat \rho) \ge  V(\rho)
\end{equation}
where the trajectory $\hat\rho(t)$ connects $\bar\rho$ to $\rho$.

Fix $t_1<t_2$, two profiles $\rho_1$, $\rho_2$ and a path $\hat\rho(t)$
in the time interval $[t_1,t_2]$ that joins $\rho_1$ to $\rho_2$:
$\hat\rho({t_1}) = \rho_1$, $\hat\rho({t_2}) = \rho_2$.
Let $H$, vanishing at the boundary, be given by the equation
\begin{equation}
\label{bf05}
\partial_t \hat\rho = \frac 12 \nabla \cdot 
\Big( D(\hat\rho)\nabla\hat\rho \Big)
- \nabla \cdot \Big( \chi(\hat\rho) \nabla H \Big) 
\end{equation}
We then claim that
\begin{eqnarray}\label{lrsq}
J_{[t_1, t_2]} (\hat\rho) &=&  V(\rho_2)\; -\; V(\rho_1) 
\nonumber\\
&&+ \frac 12 \int_{t_1}^{t_2} dt\, \Big< \nabla \Big\{ H  - \frac{\delta
V(\hat\rho)}{\delta\hat\rho} \Big\}, 
\chi(\hat\rho) \nabla \Big\{ H  - \frac{\delta
V(\hat\rho)}{\delta\hat\rho} \Big\} \Big> 
\end{eqnarray} 
Since the last term above is positive the bound \reff{lS>V}
follows from the above identity.

To prove \reff{lrsq} we note, recalling \reff{J2},  that, since $H$ is
the solution of \reff{bf05},
$$
J_{[t_1, t_2]} (\hat\rho) = \frac 12 \int_{t_1}^{t_2} dt\,
\left\langle \nabla H, \chi(\hat \rho) \nabla H\right\rangle
$$
We add and subtract in this expression $\nabla \{\delta V(\hat\rho)/\delta
\hat\rho\}$ to obtain 
\begin{eqnarray}
\label{bf04}
 J_{[t_1, t_2]} (\hat\rho) &=&  
\frac 12 \int_{t_1}^{t_2}\! dt\, \Big\langle \nabla \Big\{ H  - \frac{\delta
V(\hat\rho)}{\delta\hat\rho} \Big\} , 
\chi(\hat\rho) \nabla \Big\{ H  - \frac{\delta
V(\hat\rho)}{\delta\hat\rho} \Big\} \Big\rangle 
\nonumber \\
&&+ \int_{t_1}^{t_2} dt\, \Big\langle \nabla \frac{\delta V(\hat\rho)}
{\delta\hat\rho} , \chi(\hat\rho) \nabla H \Big\rangle 
- \frac 12 \int_{t_1}^{t_2} \!dt\, \Big\langle \nabla 
\frac{\delta V(\hat\rho)} {\delta\hat\rho} , \chi(\hat\rho) 
\nabla \frac{\delta V(\hat\rho)}{\delta\hat\rho} \Big\rangle
\phantom{merd}
\end{eqnarray}
We leave the first term unchanged and we show that the sum of the
second and third gives $V(\rho_2) - V(\rho_1)$.  Since ${\delta
V(\hat\rho)}/{\delta\hat\rho}$ vanishes at the boundary, we may integrate by
parts the second term; we get
$$
- \int_{t_1}^{t_2} \!dt\, \Big\langle \frac{\delta V(\hat\rho)}
{\delta\hat\rho}, \nabla  \cdot \Big(\chi(\hat\rho) \nabla H \Big) 
\Big\rangle
$$
By the Hamilton--Jacobi equation \reff{HJ}, the third term is equal to
$$
\frac 12 \int_{t_1}^{t_2} dt\, \Big\langle \frac{\delta V(\hat\rho)}
{\delta\hat\rho}, \nabla \cdot \Big( D(\hat\rho) \nabla \hat\rho  \Big) 
\Big\rangle
$$
Adding together the previous two expressions, we obtain that the
sum of the last terms in \reff{bf04} is equal to
$$
\int_{t_1}^{t_2} dt\, \Big< \frac{\delta V(\hat\rho)}
{\delta\hat\rho}, \Big\{ \frac{1}{2} \nabla \cdot 
\left( D(\hat\rho) \nabla \hat\rho \right)
- \nabla \cdot \left( \chi(\hat\rho) \nabla H \right) \Big\} \Big>
$$
Since $\hat\rho$ is the solution of \reff{bf05}, this expression is equal
to
$$
\int_{t_1}^{t_2} dt\, \Big< \frac{\delta V(\hat\rho)}
{\delta\hat\rho} , \partial_t\hat\rho \Big>\; =\; V(\hat\rho({t_2})) -
V(\hat\rho({t_1})) = V(\rho_2) - V(\rho_1)
$$
which proves the claim.

\ssubsection{Identification of the entropy}\label{s:2.6}
\par\noindent
In order to have a selection criterion for the solution $V(\rho)$ of
the Hamilton--Jacobi equation, we consider the partial differential
equation 
\begin{equation}
\label{bf03}
\partial_t \rho = - \frac 12 \nabla \cdot 
\Big( D(\rho) \nabla \rho \Big)
+ \nabla \cdot 
\Big( \chi(\rho) \nabla  \frac {\delta V}{\delta \rho}\Big)
\end{equation}
As in the previous Subsection we assume $V(\rho)$ is normalized so that
$V(\bar\rho)=0$ and satisfies $\delta V(\rho)/\delta \rho =0$
at the boundary.
Note that this is the adjoint hydrodynamics (\ref{adjH}) if $V$
coincides with $S$.

If $V=S$ we then have, by using the H Theorem (\ref{HT}),  that
the solution of the Cauchy problem \reff{bf03} with initial condition
$\rho$ relaxes to the stationary profile $\bar\rho$ so
that
\begin{equation}
\label{bf06}
\lim_{t\to\infty} V(\rho(t)) \; =\; V(\bar\rho)\; = 0
\end{equation}
Conversely if the above property holds, we can choose in (\ref{lrsq})
the trajectory $\hat\rho(t)=\rho(-t)$, where $\rho(t)$ solves
\reff{bf03}. We then have $H=\delta V(\hat\rho)/\delta\hat\rho$ in
(\ref{lrsq}). The last term in (\ref{lrsq}) becomes thus zero and
$V(\rho_1)$ can be made arbitrary small; therefore (\ref{lS>V}) holds
as an equality.

The above argument shows that $V=S$ if and only if (\ref{bf06}) holds.

\ssubsection{Hamiltonian interlude}
\label{s:hi}
\par\noindent
As in Section 2.2, let us interpret the large deviation functional $J$
as the action for the Lagrangian $\mathcal{L}(\rho, \partial_t \rho
)$, see (\ref{Jj}), and $J^*$ as the action for the Lagrangian
$\mathcal{L}^*(\rho, \partial_t \rho )$, see (\ref{J2*}). Let also
$\mathcal{H}(\rho ,H)$ and $\mathcal{H}^*(\rho, H)$ be the
corresponding Hamiltonians obtained as Legendre transforms, see
(\ref{lHJ}).

The time reversal relationship (\ref{J}) implies
the following relation between Lagrangians:
\begin{equation}\label{LL*}
\mathcal{L}(\rho, \partial_t \rho )=\mathcal{L}^*(\rho,-\partial_t\rho )
+\Big\langle \frac{\delta S}{\delta \rho},\partial_t\rho \Big\rangle
\end{equation}
As a consequence we obtain
\begin{equation}
\label{HH*}
\mathcal{H}(\rho, H)=
\mathcal{H}^*\Big(\rho, \,\frac{\delta S}{\delta\rho}-H \Big)
\end{equation}

Since $\mathcal{L}(\rho, \partial_t \rho )$ and
$\mathcal{L}^*(\rho,-\partial_t\rho )$ differ by a total time
derivative, see \reff{LL*}, we have the following.  Given any
$\hat{\rho}$ solution of either $\partial_t\hat{\rho}= \mf D
(\hat{\rho})$ or $\partial_t\hat{\rho}=- \mf D^* (\hat{\rho})$ then
$\hat{\rho}$ is a solution of the Euler--Lagrange equation for the
Lagrangian $\mathcal{L}$.  Likewise given any $\hat{\rho}$ solution of
either $\partial_t\hat{\rho}= \mf D^* (\hat{\rho})$ or
$\partial_t\hat{\rho}=- \mf D (\hat{\rho})$ then $\hat{\rho}$ is a
solution of the Euler--Lagrange equation for the Lagrangian
$\mathcal{L}^*$.

In the case of the quadratic functional (\ref{J2}), we have
\begin{equation}
\mathcal{L}(\rho,\partial_t \rho)=\frac 12 \left\langle 
\nabla^{-1} \left( \partial_t \hat\rho -\mf D (\rho)\right) \,,\, 
\chi(\hat\rho)^{-1} 
\nabla^{-1} \left( \partial_t \hat\rho - \mf D (\rho)\right)
\right\rangle 
\end{equation}
The momentum conjugate to $\partial_t \rho$ is
\begin{equation}\label{conjmom}
H=\frac{\delta \mathcal{L}}{\delta (\partial_t \rho)}= -
\nabla^{-1}\left( \chi (\rho )^{-1} 
\nabla^{-1} \left(\partial_t \rho -  \mf D(\rho ) \right)\right)
\end{equation}
where we recall that $\mf D(\rho )=\frac 12 \nabla \cdot (D(\rho
)\nabla \rho)$. Note that the above equation is the relationship
\reff{bf05}; as we shall see later, $H$ is the external field we have
to add to the microscopic dynamics to produce the fluctuation
$\hat\rho$.

The Hamiltonian is given by (\ref{Ham}), so that the Hamilton
equations are
\begin{equation}
\label{hameq}
\left\{
\begin{array}{lllll}
\partial_t \rho &= &
{\displaystyle \frac{\delta \mathcal{H}}{\delta H}
} &=&{\displaystyle \frac 12 \nabla \cdot \left( D(\rho )\nabla
\rho\right) -\nabla \cdot (\chi(\rho)\nabla H)} 
\\ 
\partial_t H &=&
{\displaystyle -\frac{\delta \mathcal{H}}{\delta \rho} }
&=&
{\displaystyle -\frac 12 \sum_{1\le i,j \le d} 
\left[ \chi'_{i,j}(\rho) \partial_{u_i} H \, 
\partial_{u_j}H + D_{i,j}(\rho )\partial_{u_i}\partial_{u_j}H \right] }
\end{array}
\right.
\end{equation}
where $\rho(t,u)$ satisfies the same boundary conditions as in
the hydrodynamical equation \reff{H}, $H(t,u)$ vanishes at the
boundary and $\chi'(\rho)$ is the derivative of $\chi(\rho)$ with
respect to $\rho$. 

We note that $(\bar\rho,0)$ is an equilibrium solution of \reff{hameq}
belonging to the zero energy manifold $\mathcal{H}(\rho,H)=0$. Any
solution $\rho(t)$ of the hydrodynamical equation \reff{H} corresponds
to a solution $(\rho(t), 0)$ of the Hamilton equation \reff{hameq}
which converges asymptotically, as $t\rightarrow +\infty$, to the
equilibrium point $(\bar\rho,0)$. The action $J$ evaluated on this
solution is identically zero; this corresponds to the trivial solution
$S=0$ of the Hamilton--Jacobi equation \reff{HJ} and is consistent
with the vanishing of the conjugate moment $H$.  Furthermore, if we
take the time reversal of any solution of the adjoint hydrodynamics,
i.e.\ $\partial_t \rho(t) = -\mf D^* (\rho(t))$ we find a solution of
the Hamilton equations given by $\left(\rho(t), (\delta S /
\delta\rho)(\rho(t)) \right)$ which converges asymptotically, as
$t\rightarrow -\infty$, to the equilibrium point $(\bar\rho,0)$; the
action $J$ evaluated on this solution, as a function of the final
state $\rho$, is the macroscopic entropy $S(\rho)$. Both these
trajectories live on the zero energy manifold.  Similar properties
hold for the Hamiltonian flow of $\mathcal{H}^*$.

Let us introduce the involution $\Theta$ on the phase space $(\rho,H)$
defined by
$$
\Theta (\rho,H) = \big( \rho,\, \frac{\delta S}{\delta \rho}(\rho)  -H \big) 
$$
If we denote by $\Phi_t$, resp. $\Phi_t^*$, the Hamiltonian flow of
$\mathcal H$, resp.  $\mathcal{H}^*$, 
then, by using \reff{HH*}, as easy computation shows that $\Theta$
acts as the time reversal in the sense that
\begin{equation}\label{hamtr}
\Theta \Phi_t = \Phi^*_{-t}\Theta 
\end{equation}
equivalently, in terms of the Liouville operators, we have
$$
\Theta \{f, \mathcal H\} = - \{\Theta f,\mathcal {H}^* \}
$$
where $f$ is a function on the phase space and $\{\cdot,\cdot\}$ is
the Poisson bracket.

The relationship \reff{hamtr} is non trivial also for reversible
processes, i.e.\ when $\mathcal{H}=\mathcal{H}^*$, in such a case it
tells us how to change the momentum under time reversal.  This
definition of time reversal in a Hamiltonian context 
agrees with the one given in \cite{MTR}.

\ssection{Boundary driven zero range process} 
\par\noindent 
We consider now the so called zero range process which models a
nonlinear diffusion of a lattice gas, see e.g.\ \cite{kl}.  The model
is described by a positive integer variable $\eta_{x}(\tau)$
representing the number of particles at site $x$ and time $\tau$ of a
finite subset $\Lambda_N$ of the $d$--dimensional lattice, $\Lambda_N
= \bb{Z}^d \cap N \Lambda$ where $\Lambda$ is a bounded open subset of
$\bb{R}^d$.  The particles jump with rates $g(\eta_x)$ to one of the
nearest--neighbor sites.  The function $g(k)$ is increasing and
$g(0)=0$.  We assume that our system interacts with particle
reservoirs at the boundary of $\Lambda_N$ whose activity at site $x$
is given by $\psi(x/N)$ for some given smooth strictly positive
function $\psi(u)$.  The microscopic dynamics is then defined by the
generator (see \cite{DF} for the one dimensional case)
$$
L_N = L_{N,{\rm bulk}} + L_{N, {\rm bound.}}
$$
where
\begin{equation}
\label{GEN} 
\begin{array}{lll}
\!\!\!\!\! 
{\displaystyle
L_{N,{\rm bulk}} f(\eta)} &=& {\displaystyle {\frac {1}{2}} 
\sum_{x,y\in\Lambda_N \atop |x-y|=1} 
g(\eta_x) \left[ f(\eta^{x,y}) -f (\eta)\right]}
\\
\!\!\!\!\!
{\displaystyle
L_{N,{\rm bound.}} f(\eta)} &=& {\displaystyle
{\frac {1}{2}}
\sum_{x\in\Lambda_N, y \not\in\Lambda_N 
\atop |x-y|=1} \Big\{
g(\eta_x) \left[ f(\eta^{x,-}) -f (\eta)\right]
+\psi(y/N) [f(\eta^{x,+}) - f(\eta)]  \Big\} 
}
\end{array}
\end{equation}
in which 
\begin{equation}\label{exy}
\eta^{x,y}_z  = \left\{
\begin{array}{ccl}
\eta_z &\hbox{if}& z\neq x,y \\
\eta_z -1 &\hbox{if}& z=x \\
\eta_z +1  &\hbox{if}& z=y 
\end{array}
\right.
\end{equation}
is the configuration obtained from $\eta$ when a
particle jumps from $x$ to $y$, and
\begin{equation}\label{expm}
\eta^{x,\pm}_z = 
\left\{
\begin{array}{ccl}
\eta_z &\hbox{if}& z\neq x \\
\eta_z \pm 1 &\hbox{if}& z=x 
\end{array}
\right.
\end{equation}
is the configuration where we added (resp. subtracted) one particle at
$x$. Note that, since $g(0)=0$, the number of particles cannot become
negative. 

We also remark that if $g(k)=k$ the dynamics introduced
represents simply non interacting random walks on $\Lambda_N$ (with the
appropriate boundary conditions) in terms of the occupation numbers
$\eta_x$.

\ssubsection{Invariant measure}
\par\noindent
Since the generator $L_N$ is {\em irreducible} (we can go with
positive probability from any configuration to any other), under very
general hypotheses on the function $g(k)$ there exists a unique
invariant measure.  It is however remarkable that such invariant
measure can be constructed explicitly (see \cite{DF} for the one
dimensional case).

Let $\lambda_N(x)$ be the solution of the discrete Laplace equation
with boundary condition $\psi$, namely
\begin{equation}\label{laminv}
\left\{
\begin{array}{l}
{\displaystyle 
\frac 12 \Delta_N \lambda_N (x) \equiv 
\frac 12
\sum_{y\in\bb{Z}^d \atop |x-y|=1} [\lambda_N (y) -\lambda_N(x)] = 0
\quad \textrm{for any}\;\; x\in\Lambda_N 
}
\\
{\displaystyle 
\lambda_N(x) = \psi(x/N)
\quad \textrm{for any}\;\; x\not\in\Lambda_N \;\;\textrm{such that}
\;\; \exists y\in\Lambda_N \textrm{ for which } |x-y|=1}
\end{array}
\right.
\end{equation}
The invariant measure $\mu_N$ is the grand--canonical measure
$\mu_N=\prod_{x\in\Lambda_N} \mu_{x,N}$ obtained by
taking the product of the marginal distributions
\begin{equation}
\mu_{x,N}
(\eta_x = k) = \frac {1}{Z(\lambda_N(x))} \; 
{\frac {\lambda_N^k (x)}{g(1)\cdots g(k)}} 
\label{INV}
\end{equation}
where 
\begin{equation}
\label{Z=}
Z(\varphi) = 1 + \sum_{k=1}^{\infty}{\frac {\varphi^k}{g(1)
\cdots g(k)}}
\end{equation}
is the normalization constant.

The fact that $\mu_N$ is an invariant measure can be verified by
showing that 
$$
\sum_{\eta} \mu_N(\eta) L_N f (\eta) = 0
$$
for any bounded observable $f$. The above identity can be easily checked
taking into account that, since $\lambda_N$ solves \reff{laminv}, 
it is an harmonic function; in particular we have
\begin{equation}\label{arm}
\sum_{x\in\Lambda_N, y \not\in\Lambda_N 
\atop |x-y|=1} \left[ \lambda_N(y) -\lambda_N(x)\right] = 0
\end{equation}

We emphasize that, if $\psi$ is not constant, the generator $L_N$
is not self--adjoint with respect to the invariant measure so that the
process is different from its time reversal and detailed balance does
not hold.

\ssubsection{Hydrodynamic limit}
\par\noindent
Let us introduce now the macroscopic time $t= \tau /N^2$ and space 
$u = x/N$. For $u\in\Lambda$, $t\ge 0$, we introduce the empirical
density as
\begin{equation}
\rho_N(t,u) = \frac{1}{N^d} \sum_{x\in\Lambda_N} \eta_x(N^2 t) 
 \: \delta \left(u - \frac xN \right)
\label{ED}
\end{equation}
where $\delta$ denotes the Dirac function. Note that 
$$
\int_\Lambda\!du\: \rho_N(t,u) = \frac{1}{N^d} \sum_{x\in\Lambda_N} 
\eta_x({N^2 t }) 
$$
is the average density of particles at (macroscopic) time $t$.

Let  $G(u)$, $u\in\Lambda$ be a smooth function and consider
$$
\langle \rho_N(t), G \rangle = \int_\Lambda\!du\: \rho_N(t,u) \, G(u)
$$
To compute the time evolution of $\langle \rho_N(t), G \rangle$ 
we first note that, according to the general theory of Markov
processes, see e.g.\ \cite[Ch. X]{FE}, we have
\begin{equation}
\frac{d}{dt}  {\bb E}^N_\eta \left( \langle \rho_N(t), G \rangle \right) 
=   {\bb E}^N_\eta \left(  N^2 L_N \langle\rho_N(t), G \rangle \right) 
\end{equation}
where ${\bb E}^N_\eta$ denotes the expectation with respect to 
the microscopic process with initial configuration $\eta$.

Let us assume that $G$ has compact support $K\subset \Lambda$, so that
only $L_{N,{\rm bulk}}$ gives a non zero contribution; by
summing by parts, we get
\begin{eqnarray}
\label{Lr}
N^2 L_N \langle \rho_N(t), G  \rangle &=&
\frac 12 \frac {1}{N^d} \sum_{x\in\Lambda_N} g(\eta_x({N^2 t }))
N^2 \Delta_N G (x/N)
\nonumber\\
&\approx&
\frac 12 \frac {1}{N^d} \sum_{x\in\Lambda_N} g(\eta_x({N^2 t }))
\Delta G (x/N) 
\end{eqnarray}
where we recall that $\Delta_N$ denotes the discrete Laplacian.

At this point we face the main problem in the hydrodynamical limit:
equation \reff{Lr} is not closed in $\rho_N(t)$ (its r.h.s.\ is not a
function of $\rho_N(t)$). In order to derive the macroscopic
hydrodynamic equation we need to express $g(\eta_x({N^2 t }))$ in
terms of the empirical density $\rho_N(t)$. This will be done by
assuming a ``local equilibrium'' state (which can be rigorously
justified in this context).  Let us consider a microscopic site $x$
which is at distance $O(N)$ from the boundary and introduce a volume
$Q$, centered at $x$, which is very large in microscopic units, but
still infinitesimal at the macroscopic level. The time evolution in
$Q$ is essentially given only by the bulk dynamics $L_{N,{\rm bulk}}$;
since the total number of particles in $Q$ changes only via boundary
effects and we are looking at what happened after $O(N^2)$ microscopic
time units, the system in $Q$ has relaxed to the canonical state
corresponding to the density $\rho_N(t,x/N)$.

Let us construct first the grand--canonical measure in $Q$
with constant activity $\varphi>0$, namely the product measure
$\mu^\varphi_Q =\prod_{x\in Q}\mu^\varphi_x$ with marginal given by
$$
\mu^\varphi_x (\eta_x=k) = \frac{1}{Z(\varphi)} 
{\frac {\varphi^k}{g(1)\cdots g(k)}} 
$$
where $Z(\varphi)$, which has been defined in \reff{Z=}, is the
normalization constant. 
Let now $\nu_{Q}^\alpha$ be the associated canonical measure
at density $\alpha$, i.e.\
$$
\nu_Q^\alpha (\eta) = \mu^\varphi_Q 
\left( \eta \left| \sum_{x\in Q} \eta_x
=\alpha |Q| \right. \right)
$$ 
We introduce a function $\phi(\alpha)$ by 
\begin{equation}\label{f-1}
\phi(\alpha) =\lim_{Q\uparrow \bb Z^d} 
E^{\nu_Q^\alpha} \big( g (\eta_x ) \big)
\end{equation}
where we recall that $E^{\nu_Q^\alpha}$ denotes the expectation with
respect to the probability $\nu_Q^\alpha$.

According to the previous discussion, the system in the volume $Q$ is
well approximated by a canonical state with density $\rho_N(t,x/N)$; 
we can thus replace, for $N$ large,
$g(\eta_x({N^2t}))$ on the r.h.s.\ of \reff{Lr} by
$\phi(\rho_N (t,x/N))$ thus obtaining
\begin{equation}\label{drho}
\frac {d}{dt} {\bb E}^N_\eta \left( \langle \rho_N(t), G \rangle   \right) 
\approx  \frac 12 {\bb E}^N_\eta 
\left( \langle \phi(\rho_N(t)), \Delta G \rangle \right)
\end{equation}

To see which are the boundary conditions satisfied by $\rho_N(t)$ we
need to consider the boundary dynamics, $L_{N,{\rm bound.}}$. In
contrast to the bulk dynamics, this is a non conservative,
Glauber--like, dynamics. Since we are looking after $O(N^2)$
microscopic time units the density at the boundary has well
thermalized to its equilibrium value which impose
\begin{equation}
\label{rbound}
{\bb E}^N_\eta \left( \phi\left( \rho_N(t,u) \right)  \right) 
\approx \psi(u) \quad\quad u\in\partial \Lambda
\end{equation}
where the function $\phi$ has been defined above.

Assume the initial configuration $\eta$ of the process is such that
for any smooth function $G$ we have
\begin{equation}\label{0g}
\lim_{N\to\infty}\langle \rho_N(0), G \rangle 
= \lim_{N\to\infty} \frac {1}{N^d} 
\sum_{x\in\Lambda_N} \eta_x G(x/N)
=\int_\Lambda\! du \: \gamma(u) G(u)
\end{equation}
for some function $\gamma$.  By the law of large numbers, $\rho_N(t)$
becomes a deterministic function in the limit $N\to\infty$, so that we
can eliminate the expectation values in \reff{drho} and \reff{rbound}.
In conclusion we have obtained, for any smooth function $G$, that
$$
\lim_{N\to\infty}  
\langle \rho_N(t), G \rangle = \int_\Lambda\! du \: \rho(t,u) G(u)
$$
where the convergence is in probability.
Recalling \reff{drho}, the limiting density $\rho=\rho(t,u)$ solves 
\begin{equation}\label{H0R}
\left\{
\begin{array}{l} 
{\displaystyle
\partial_t \rho (t,u) = \frac 12 \Delta \phi( \rho (t,u) ) \quad u \in \Lambda
}
\\
{\displaystyle \phi\left( \rho(t,u)  \right) = \psi(u) \quad u \in
\partial\Lambda} 
\\
{\displaystyle \rho(0,u)   = \gamma(u) } 
\end{array}
\right.
\end{equation}
which is the hydrodynamic equation for the boundary driven zero range
process.

We finally show that, by the equivalence of ensembles, we can express
the function $\alpha\mapsto \phi(\alpha)$ introduced in \reff{f-1}, in
a more convenient way, in terms of the grand--canonical measure
$\mu_Q^\varphi$. By exploiting the product structure of
$\mu_Q^\varphi$ and choosing the activity $\varphi(\alpha)$ 
so that $\mu_x^{\varphi(\alpha)} (\eta_x) =\alpha$, we have 
$$
\phi(\alpha) = E^{\mu_x^{\varphi(\alpha)}} \big( g(\eta_x) \big) 
$$
A straightforward computation shows now that
$\phi(\alpha)=\varphi(\alpha)$ so that the function
$\alpha\mapsto\phi(\alpha)$ is the inverse of the function $\varphi
\mapsto R(\varphi)$ given by
\begin{equation}\label{Rf}
R(\varphi) = E^{\mu_x^\varphi} (\eta_x) = \varphi 
\, \frac {Z'(\varphi)}{Z(\varphi)}
\end{equation}
i.e.\ $R(\varphi)$ is the equilibrium density corresponding to
the activity $\varphi$.  From the assumptions on $g$ it follows that
$R(\varphi)$ is strictly increasing.

\ssubsection{Dynamical large deviations}
\par\noindent
In order to compute the probability of deviation from the typical
behavior described by equation \reff{H0R}, namely the action
functional $J(\hat\rho)$, we follow the classical strategy in large
deviation theory: we need ``only'' to consider a perturbation of the
system which makes the deviation $\hat\rho$ the typical behavior and
write the probability in the unperturbed system in terms of the
perturbed one. From this computation we shall extract, asymptotically
in $N$, the factor $\exp\{-N^d J(\hat\rho)\}$.

We consider the zero range process in a (space time dependent)
external field $H(t,u)$ which is a smooth function of the macroscopic
variables vanishing outside $\Lambda$, i.e.\ $H(t,u)=0$ for
$u\not\in\Lambda$, $t\ge 0$.  The perturbed dynamics is specified by
the time dependent generator $L_{N,\tau}^H = 
L_{N,\tau,{\rm bulk}}^H+L_{N,\tau,{\rm bound.}}^H$ where
\begin{equation}\label{pgen}
\begin{array}{lll}
{\displaystyle 
L_{N,\tau,{\rm bulk}}^H f (\eta)} &=& {\displaystyle \frac 12 
\sum_{\scriptstyle x,y\in \Lambda_N \atop \scriptstyle |x-y|=1 } 
g(\eta_x) e^{H(\tau/N^2,y/N)-H(\tau/N^2,x/N)} 
\left[f(\eta^{x,y}) - f(\eta) \right]} 
\\
{\displaystyle 
L_{N,\tau, {\rm bound.}}^H f(\eta)} &=&{\displaystyle 
{\frac {1}{2}}
\sum_{x\in\Lambda_N, y \not\in\Lambda_N 
\atop |x-y|=1} \Big\{
g(\eta_x) e^{H(\tau/N^2,y/N)-H(\tau/N^2,x/N)}  
\left[ f(\eta^{x,-}) -f (\eta)\right]
}\\
&&\;\;\;\; {\displaystyle
+\, \psi(y/N) e^{H(\tau/N^2,x/N)-H(\tau/N^2,y/N)}  
[f(\eta^{x,+}) - f(\eta)]  \Big\} } 
\end{array}
\end{equation}
The interpretation of the perturbed dynamics is the following: in the
macroscopic scale, we simply introduced a small space--time dependent
drift $N^{-1} \nabla H(t,u)$ in the motion of the particles.

Assuming the initial configuration $\eta$ is associated to a density
profile $\gamma$ in the sense of \reff{0g}, by similar computations as
the ones given for $H=0$, we get that the hydrodynamic equation for
the perturbed system is
\begin{equation}\label{H0RP}
\left\{
\begin{array}{l} 
{\displaystyle
\partial_t \hat\rho (t,u) = \frac 12 \Delta \phi( \hat\rho (t,u) ) - 
\nabla\cdot \left( \phi( \hat\rho (t,u) ) \nabla H(t,u) \right)
\quad u \in \Lambda
}
\\
{\displaystyle \phi\left( \hat\rho(t,u)  \right) = \psi(u) \quad u \in
\partial\Lambda} 
\\
{\displaystyle \hat\rho(0,u)  = \gamma(u) } 
\end{array}
\right.
\end{equation}
If we regard $\hat\rho(t,u)$ as given and consider $H(t,u)$ as the
unknown, the above equation tells us which is the perturbation for which
$\hat\rho(t,u)$ is the typical behavior.  Note \reff{H0RP} is
precisely the relationship \reff{conjmom} between $\partial_t\hat\rho$
and the conjugate momentum $H$.

Let us denote by $\bb P^N_\eta$, resp.  $\bb P_\eta^{N,H}$, the
probability for the unperturbed, resp. perturbed, process with initial
configuration $\eta$. We then have
\begin{equation}\label{RN}
\frac{ d \bb P_\eta^{N,H} \left( \eta(N^2t), \: t\in [0,T] \right) }
{ d \bb P_\eta^{N} \left( \eta(N^2 t ), \: t\in[0,T] \right)}
= \exp\left\{ \mathcal{J}_{[0,T]}^N (\eta (\cdot),H) \right\}
\end{equation}
where
\begin{eqnarray}\label{strJ}
\mathcal{J}_{[0,T]}^N (\eta (\cdot),H)
&=& \sum_{x\in\Lambda_N} \left[ 
H(T, x/N) \eta_x(N^2T) - H(0, x/N) \eta_x(0) \right] 
\nonumber \\
&&\;\;
- \int_0^T \!dt \: \sum_{x\in\Lambda_N} \partial_t H(t, x/N) \eta_x(N^2t)
\nonumber\\
&&\;\;
- \frac{N^2}{2}  \int_0^{T} \!dt \: 
\sum_{\scriptstyle x\in \Lambda_N, y\in{\bb Z}^d
\atop \scriptstyle |x-y|=1}  
g(\eta_x(N^2 t)) \left[ e^{H(t,y/N)-H(t,x/N)} - 1 \right]
\nonumber\\
&&\;\;
-\frac{N^2}{2}  \int_0^{T} \!dt \: 
\sum_{\scriptstyle x\in \Lambda_N, y\not\in\Lambda_N
\atop \scriptstyle |x-y|=1}  
\psi(y/N) \left[ e^{H(t,x/N)-H(t,y/N)} - 1 \right]
\phantom{merd}
\end{eqnarray}
See Appendix \ref{appa} for a derivation of the above formula 
(another proof can be found in \cite[Prop. A1.7.3]{kl}).

With the help of equation \reff{RN} we can write the probability
that $\rho_N(t)$ is close to $\hat\rho(t)$ for the unperturbed system as
follows 
\begin{equation}\label{perunper}
{\bb P}_\eta^N \left( \rho_N(t) \sim \hat\rho(t) , \: t\in[0,T] \right)
= {\bb E}_\eta^{N,H} \left( e^{-\mathcal{J}_{[0,T]}^N (\eta (\cdot),H)}
\id_{\rho_N \sim \hat\rho } \right)
\end{equation}
where $ {\bb E}_\eta^{N,H}$ denotes the expectation with respect to the
perturbed process. 
Using the explicit expression \reff{strJ}, by Taylor expansion, we get
\begin{eqnarray}
\mathcal{J}_{[0,T]}^N (\eta (\cdot),H)
&\approx &  N^d \left\{ \langle \rho_N(T), H(T) \rangle -
\langle \rho_N(0), H(0) \rangle
- \int_0^T\!dt \: \langle \rho_N(t), \partial_t H(t) \rangle
\right.
\nonumber\\
&&
\left.
- \frac 12 \int_0^{T} \!dt \: 
\frac {1}{N^d} \sum_{x\in\Lambda_N} g(\eta_x(N^2t))
\left[ \Delta H (t,x/N) + \left| \nabla H (t,x/N) \right|^2 \right] 
\right.
\nonumber\\
&&
\left.
+ \frac 12 \int_0^{T} \!dt \: \int_{\partial\Lambda} \!du \: 
\psi(u) \partial_{\hat n} H(t,u)
\right\}
\end{eqnarray}
where $\partial_{\hat n} H(t,u)$ is the normal derivative of $ H(t,u)$
($\hat n$ being the outward normal to $\Lambda$).

If $\eta(N^2t)$ is a typical trajectory for the perturbed process, by
the same argument given in derivation of the hydrodynamical equation,
we can replace $g(\eta_x(N^2t))$ above by $\phi
(\rho_N(t,x/N))$. Since $\rho_N(t)$ is close to $\hat\rho(t)$, and
$$
\lim_{N\to\infty} 
{\bb P}_\eta^{N,H} \left(\rho_N(t) \sim  \hat\rho(t) , \; t\in [0,T] \right) 
= 1 
$$ from \reff{perunper} we get 
\begin{equation}
{\bb P}_\eta^N \left( \rho_N(t) \sim \hat\rho(t) , \: t\in[0,T] \right)
\approx \exp\left\{ - N^d J_{[0,T]} (\hat\rho) \right\}
\end{equation}
where
\begin{eqnarray}
&&J_{[0,T]} (\hat\rho)
=  \langle \hat\rho(T), H(T) \rangle -
\langle \hat\rho(0), H(0) \rangle
- \int_0^T\!dt \: \langle \hat\rho(t), \partial_t H(t) \rangle
\nonumber \\
&&\;\;\;
- \frac 12 \int_0^{T} \!dt \: 
\left\langle \phi(\hat\rho(t)), \Delta H (t) + \left| \nabla H (t)
\right|^2 \right\rangle  
+ \frac 12 \int_0^{T} \!dt \: \int_{\partial\Lambda} \!du \: 
\psi(u) \partial_{\hat n} H(t,u) \phantom{merd}
\end{eqnarray}
Recalling that $H$ and $\hat\rho$ are related by \reff{H0RP}, we
finally get, after an integration by parts (recall that $H(t,u)$
vanishes at the boundary $\partial\Lambda$)
\begin{equation}\label{J0R}
J_{[0,T]} (\hat\rho) = \frac 12 \int_0^{T} \!dt \: 
\left\langle \phi(\hat\rho(t)) , 
\left| \nabla H (t) \right|^2 \right\rangle
\end{equation}
The action functional $J$ is defined to be infinite if $\hat\rho$ does
not satisfy the boundary conditions in \reff{H0RP}.  From \reff{J0R}
and \reff{H0RP} we see that $J_{[0,T]}$ is of the form \reff{J2} with
$D_{i,j}(\hat\rho)= \phi'(\hat\rho) \delta_{i,j}$ and
$\chi_{i,j}(\hat\rho)= \phi(\hat\rho) \delta_{i,j}$.

The rigorous derivation of the action functional $J$ requires some
difficult estimates.  In fact, while in the proof of the hydrodynamic
limit it is enough to show that we can replace $g(\eta_x(N^2t))$ by
$\phi(\rho_N(t,x/N))$ with an error vanishing as $N\to\infty$, in the
proof of the large deviations we need such an error to be
$o(e^{-N^d})$. This can be proven by the so called super exponential
estimate, see \cite{kl,kov}, which is the key point in the rigorous
approach.

\ssubsection{Macroscopic entropy and adjoint hydrodynamics}
\par\noindent From the expression \reff{J0R} for $J$ it follows that the
Hamilton--Jacobi equation \reff{HJ} for the boundary driven zero range
process is
\begin{equation}\label{hj0r}
\left\langle \nabla \frac{\delta S}{\delta \rho}, \phi(\rho)  
\nabla \frac{\delta S}{\delta \rho} \right\rangle
+\left\langle  \frac{\delta S}{\delta \rho}, 
\Delta \phi(\rho) \right\rangle =0
\end{equation}

Let us consider the functional
\begin{equation}
S(\rho)=
\int_\Lambda \!du \: 
\left[ \rho(u) \log \frac {\phi(\rho(u))}{\lambda(u)}  
- \log \frac{Z(\phi(\rho(u)))}{Z(\lambda(u))}
\right]
\label{E}
\end{equation}
where $Z(\varphi)$ has been defined in \reff{Z=} and $\lambda(u)$ is
the stationary activity profile, namely $\lambda(u)= \phi
(\bar\rho(u))$ where $\bar\rho$ is the stationary solution of the
hydrodynamic equation \reff{H0R}. Note that $\lambda$ is also the
macroscopic limit of $\lambda_N$, solution of \reff{laminv}.  
By using that $\varphi\mapsto R(\varphi)$ given in \reff{Rf} is the
inverse function of $\rho\mapsto\phi(\rho)$, we get
\begin{equation}\label{dS}
\frac{\delta S(\rho)}{\delta \rho(u)} 
= \log \phi(\rho(u)) - \log \lambda(u) 
\end{equation}
We remark that the functional $S$ given in \reff{E} is uniquely
characterized by \reff{dS} once we impose the normalization
$S(\bar\rho)=0$. 

An easy computation shows that the functional $S$ given in \reff{E}
solves the Hamilton--Jacobi equation \reff{hj0r}.  Recalling that
$\phi(\rho(u)) =\lambda(u) = \psi(u)$ for $u\in\partial\Lambda$ we
have indeed
\begin{eqnarray}\label{vhj0r}
&&
\left\langle \nabla \log\frac{\phi(\rho)}{\lambda}, \phi(\rho)
\nabla \log\frac{\phi(\rho)}{\lambda}\right\rangle
+\left\langle \log\frac{\phi(\rho)}{\lambda},
\Delta \phi(\rho)\right\rangle
\nonumber\\
&&\;\;\;= -\left\langle \nabla \log\frac{\phi(\rho)}{\lambda}, 
\phi(\rho) \frac {\nabla\lambda}{\lambda}\right\rangle
\nonumber\\
&&\;\;\;= 
\left\langle \phi(\rho), \frac {\left|\nabla\lambda\right|^2}{\lambda^2}
\right\rangle
- \left\langle \nabla \left[\phi(\rho) -\lambda\right] , 
\frac {\nabla\lambda}{\lambda} \right\rangle
-  \left\langle \nabla\lambda ,
\frac {\nabla\lambda}{\lambda}
\right\rangle
\nonumber\\
&&\;\;\;= 
\left\langle \phi(\rho), \frac {\left|\nabla\lambda\right|^2}{\lambda^2}
\right\rangle
-  \left\langle \nabla\lambda ,
\frac {\nabla\lambda}{\lambda}
\right\rangle
+ \left\langle \left[\phi(\rho) -\lambda\right],  
\nabla \cdot  \frac {\nabla\lambda}{\lambda} \right\rangle
=0
\end{eqnarray}
since $\Delta\lambda (u)=0$ for $u\in\Lambda$.

{}From the fluctuation--dissipation relationship \reff{gfd} we get that the
adjoint hydrodynamic equation for the boundary driven zero range
process is 
\begin{equation}\label{adjH0R}
\left\{
\begin{array}{l} 
{\displaystyle
\partial_t \rho^* (t,u) = 
\frac 12 \Delta \phi( \rho^* (t,u) )
-\nabla \cdot\left( \phi( \rho^* (t,u) ) \nabla\log\lambda(u) \right)
} \quad u \in \Lambda
\\
{\displaystyle \phi\left( \rho^*(t,u)  \right) = \psi(u) \quad u \in
\partial\Lambda} 
\\
{\displaystyle \rho^*(0,u)  = \gamma(u) } 
\end{array}
\right.
\end{equation}
Recalling that $\lambda(u)= \phi(\bar\rho(u))$, the density
profile $\bar\rho$ is also a stationary solution of \reff{adjH0R}.  
Since $\phi'(\alpha) >0$, the right hand side of \reff{adjH0R} is
dissipative; therefore we have that $\rho^*(t)\to \bar\rho$ as
$t\to\infty$; so that we meet the criterion \reff{bf06}.

\medskip
Since in this model we know explicitly the invariant measure
$\mu_N$ we can check whether the predictions \reff{E} on the macroscopic
entropy and \reff{adjH0R} on the adjoint hydrodynamics of the general
theory are correct. 

Given a smooth function $h(u)$ let us introduce the pressure $G(h)$
corresponding to the chemical potential profile $h$ as
\begin{eqnarray}\label{p0R}
G(h)&=&\lim_{N\rightarrow\infty} \frac {1}{N^d} 
\log E^{\mu_N} 
\left( \exp\left\{ N^d \langle \rho_N, h \rangle\right\} \right)
\nonumber\\
&=&\lim_{N\rightarrow\infty} \frac {1}{N^d} 
\log \sum_{\eta} \mu_N(\eta) 
\exp\left\{ \sum_{x\in\Lambda_N} h(x/N) \eta_x \right\}
\nonumber\\
&=&\lim_{N\rightarrow\infty}\frac {1}{N^d} \sum_{x\in\Lambda_N} 
\left[
\log Z\left( \lambda_N(x)e^{h(x/N)}  \right) 
-  \log Z\left( \lambda_N(x)  \right)  \right]
\nonumber\\
&=& \int_\Lambda\!du \: \left[\log Z\left( \lambda(u)e^{h(u)} \right) 
-  \log Z\left( \lambda(u)  \right)  \right]
\end{eqnarray}
where $Z(\varphi)$ has been defined in \reff{Z=} and $\lambda_N(x)$
is the solution of \reff{laminv}. 

By a standard computation due to Cramer we have that the Legendre
transform of the pressure $G(h)$, i.e.\ the macroscopic entropy, is
the rate functional for the asymptotic probability of large deviations
of the density profile in the invariant measure $\mu_N$.
Let us in fact introduce a perturbed measure $\mu_N^{h}$ by
$$
\mu_N^{h} (\eta) = \prod_{x\in\Lambda_N}  
\frac{Z\left(\lambda_N(x)\right)}{Z\left(\lambda_N(x) e^{h(x/N)} \right)} 
\: \exp\{ h(x/N) \eta_x \}  \:\mu_{x,N} (\eta_x) 
$$
and, for a fixed density profile $\rho(u)$, choose the chemical
potential profile $h$ so that 
\begin{equation}\label{choseh}
E^{\mu_N^{h}} (\eta_x) = \rho(x/N)
\end{equation}
namely $h(x/N)= \log [ \phi (\rho(x/N)) / \lambda_N(x) ]$.
We then have 
\begin{eqnarray}\label{eqcomp}
\mu_N( \rho_N \sim \rho ) &=& 
E^{\mu_N^{h}} \left( 
e^{-\sum_{x\in\Lambda_N} \left[ h(x/N)\eta_x 
- \log Z\left(\lambda_N(x) e^{h(x/N)} \right)      
+ \log Z\left(\lambda_N(x)\right) \right] } \id_{\rho_N \sim \rho} 
\right)
\nonumber\\
&\approx& E^{\mu_N^{h}} \left( 
e^{- N^d \left[ \langle \rho_N, h\rangle -  G(h) \right]}
\id_{\rho_N \sim \rho} 
\right) \approx e^{-N^d  \left[ \langle \rho, h\rangle -  G(h) \right] }
\end{eqnarray}
since, by the law of large numbers,
$\lim_{N\to\infty} \mu_N^{h}( \rho_N \sim \rho ) =1$.  From
\reff{choseh} we have that $\delta G / \delta h = \rho$.  We
therefore have obtained precisely that $S$ is the Legendre transform
of $G$. A computation, which is left to the reader, shows now 
that the Legendre transform of the right hand side of \reff{p0R}
gives indeed the functional $S(\rho)$ defined in \reff{E}.

We want to stress a main difference between the macroscopic
computation \reff{vhj0r} and the microscopic one just given. While the
former depends on the action functional $J$, which involves only
macroscopic quantities, the latter depends on the explicit expression
of the invariant measure $\mu_N$. In particular the macroscopic
computation is independent of the specific way the interaction with
reservoirs is modeled (provided of course the functional $J$ is not
affected).

\smallskip
We now discuss the adjoint hydrodynamics from a microscopic point of
view. Since the invariant measure $\mu_N$ is explicitly known we
can obtain the adjoint generator $L_N^*$ which is defined by the identity
\reff{L*}.
Recalling that $\lambda_N$ solves \reff{laminv} and that \reff{arm}
holds, we have that $L_N^* = L_{N,{\rm bulk}}^* + L_{N, {\rm
bound.}}^*$ where
\begin{eqnarray}
\label{GEN*}
L_{N,{\rm bulk}}^* f(\eta) &=&  {\frac {1}{2}} 
\sum_{x,y\in\Lambda_N \atop |x-y|=1} 
g(\eta_x) \frac{\lambda_N(y)}{\lambda_N(x)} 
\left[ f(\eta^{x,y}) -f (\eta)\right]
\\
L_{N,{\rm bound.}}^* f(\eta) &=&
{\frac {1}{2}}
\sum_{x\in\Lambda_N, y \not\in\Lambda_N 
\atop |x-y|=1} \Big\{
g(\eta_x)  \frac{\psi(y/N)}{\lambda_N(x)} 
\left[ f(\eta^{x,-}) -f (\eta)\right]
+\lambda_N(x) [f(\eta^{x,+}) - f(\eta)]  \Big\} 
\nonumber
\end{eqnarray}
Notice that the form of (\ref{GEN*}) is the same as (\ref{GEN}) with
the rates modified in such a way to invert the particle flux.  The
generator $L_N^*$ can also be interpreted as the original system
perturbed by a time independent external field $H(t,x/N)=\log
\lambda_N(x)$, compare \reff{pgen} to \reff{GEN*}. In particular we
have that the adjoint hydrodynamic equation is indeed given by
\reff{adjH0R}.

By repeating the same argument given in Section 3.3, it is not
difficult to show that the action functional $J^*$ describing the
probability of large deviations from the hydrodynamic behavior for the
adjoint process is given by
\begin{equation}\label{J0R*}
J^*_{[0,T]} (\hat\rho) = \frac 12 \int_0^{T} \!dt \: 
\left\langle \phi(\hat\rho(t)) , 
\left| \nabla K(t) \right|^2 \right\rangle
\end{equation}
where $K(t)=K(t,u)$ is to be obtained from $\hat\rho$ by solving the
equation 
\begin{equation}\label{H0RP*}
\left\{
\begin{array}{l} 
{\displaystyle
\partial_t \hat\rho (t,u) = \frac 12 \Delta \phi( \hat\rho (t,u) ) - 
\nabla\cdot \left( \phi( \hat\rho (t,u) ) 
\nabla \left[ \log\lambda(u) + K(t,u)\right] \right)
\quad u \in \Lambda
}
\\
{\displaystyle \phi\left( \hat\rho(t,u)  \right) = \psi(u) \quad u \in
\partial\Lambda} 
\\
{\displaystyle \hat\rho(0,u)  = \gamma(u) } 
\end{array}
\right.
\end{equation}
A computation now allows us to check that the identity
\reff{J}, which has been obtained only by a time symmetry argument,
holds for the boundary driven zero range process.

\ssection{Boundary driven simple exclusion process}
\par\noindent
The simple exclusion process is a model of a lattice gas with an
exclusion principle: a particle can move to a neighboring site, with
rate $1/2$ for each side, only if this is empty. Let, as in the
previous Section, $\Lambda_N = \bb{Z}^d \cap N \Lambda$ and denote by
$\eta_{x}(\tau) \in \{0,1\}$ the number of particles at the site $x$
at (microscopic) time $\tau$. The system is in contact with particle
reservoirs at the boundary of $\Lambda_N$ whose activity at site
$x$ is given by $\psi(x/N)$ for some given strictly positive smooth
function $\psi(u)$.

The microscopic dynamics is defined by the generator 
$L_N = L_{N,{\rm bulk}} + L_{N, {\rm bound.}}$
where
\begin{eqnarray}
\label{GENsep} 
L_{N,{\rm bulk}} f(\eta) &=& {\frac {1}{2}} 
\sum_{x,y\in\Lambda_N \atop |x-y|=1} 
\eta_x (1-\eta_y) \left[ f(\eta^{x,y}) -f (\eta)\right]
\\
L_{N,{\rm bound.}} f(\eta) &=& 
{\frac {1}{2}}
\sum_{x\in\Lambda_N, y \not\in\Lambda_N 
\atop |x-y|=1} \Big\{
\eta_x \left[ f(\eta^{x,-}) - f (\eta)\right]
+ \psi(y/N) (1-\eta_x)  \left[ f(\eta^{x,+}) -f (\eta)\right]
\Big\} 
\nonumber
\end{eqnarray}
where $\eta^{x,y}$ and $\eta^{x,\pm}$ have been defined in 
\reff{exy} and \reff{expm}.

In contrast to the zero range model the invariant measure $\mu_N$
is not known explicitly; we shall see that it carries long range
correlations making the entropy non local.

\ssubsection{Hydrodynamic equation and dynamical large deviations}
\par\noindent
The hydrodynamic equation for the simple exclusion process can be
derived by the same argument given for the zero range process; in fact
it is easier in this case because a computation analogous to \reff{Lr}
leads directly to a closed equation in the empirical density.  We find
that the limiting density evolves according to the linear heat
equation
\begin{equation}\label{HE}
\left\{
\begin{array}{l} 
{\displaystyle
\partial_t \rho (t,u) = \frac 12 \Delta \rho (t,u)  \quad u \in \Lambda
}
\\
{\displaystyle \rho(t,u)   = \frac{\psi(u)}{1+\psi(u)} \quad u \in
\partial\Lambda} 
\\
{\displaystyle \rho(0,u)   = \gamma(u) } 
\end{array}
\right.
\end{equation}
where $\gamma$ is the initial density profile, associated to the
initial microscopic configuration $\eta$ in the sense \reff{0g}.
In this case the density of particles $\rho$ takes value in $[0,1]$.
The hydrodynamic limit for more general boundary driven models has
been discussed in \cite{els1,els2}.
As in the previous Section we shall denote by $\bar\rho=\bar\rho(u)$
the unique stationary solution of \reff{HE}.

\smallskip
The action functional $J$ describing the probability of large
deviations from the hydrodynamic behavior can be obtained as for the
zero range process. In this case the perturbed dynamics is defined by
the time dependent generator 
$L_{N,\tau}^H = L_{N,\tau,{\rm bulk}}^H+L_{N,\tau,{\rm bound.}}^H$ where
\begin{equation}\label{pgensep}
\begin{array}{lll}
{\displaystyle 
L_{N,\tau,{\rm bulk}}^H f (\eta)} &=& {\displaystyle \frac 12 
\sum_{\scriptstyle x,y\in \Lambda_N \atop \scriptstyle |x-y|=1 } 
\eta_x (1-\eta_y)  e^{ H(\tau/N^2,y/N)-H(\tau/N^2,x/N) } 
\left[f(\eta^{x,y}) - f(\eta) \right]} 
\\
{\displaystyle
L_{N,\tau,{\rm bound.}}^H  f(\eta)} &=& {\displaystyle
{\frac {1}{2}}
\sum_{x\in\Lambda_N, y \not\in\Lambda_N 
\atop |x-y|=1} \Big\{
\eta_x e^{ H(\tau/N^2,y/N)-H(\tau/N^2,x/N) } 
\left[ f(\eta^{x,-}) - f (\eta)\right]}
\\
&&{\displaystyle \!\!\!\!\!\!
+ \psi(y/N) (1-\eta_x)   e^{ H(\tau/N^2,x/N)-H(\tau/N^2,y/N) } 
\left[ f(\eta^{x,+}) -f (\eta)\right]
\Big\} 
}
\end{array}
\end{equation}
and the external field $H(t,u)$ vanishes for $u\not\in\Lambda$.

The hydrodynamic equation for the perturbed dynamics is then given by
\begin{equation}\label{HsepP}
\left\{
\begin{array}{l} 
{\displaystyle
\partial_t \hat\rho (t,u) = \frac 12 \Delta \hat\rho (t,u)  - 
\nabla\cdot \Big\{ \hat\rho (t,u)  \left[ 1- \hat\rho (t,u)\right] 
\nabla H(t,u) \Big\}
\quad u \in \Lambda
}
\\
{\displaystyle \hat\rho(t,u) = \frac{\psi(u)}{1+\psi(u)} \quad u \in
\partial\Lambda} 
\\
{\displaystyle \hat\rho(0,u)  = \gamma(u) } 
\end{array}
\right.
\end{equation}
For the simple exclusion process the functional
$\mathcal{J}_{[0,T]}^N$ defined in \reff{RN} is given by 
\begin{eqnarray}\label{strJsep}
&&\mathcal{J}_{[0,T]}^N (\eta (\cdot),H)
= \sum_{x\in\Lambda_N} \left[ 
H(T, x/N) \eta_x(N^2T) - H(0, x/N) \eta_x(0) \right] 
\nonumber\\
&&\;\;\;\;\; 
- \int_0^T \!dt \: \sum_{x\in\Lambda_N} \partial_t H(t, x/N) \eta_x(N^2t)
\nonumber\\
&&\;\;\;\;\; - \frac{N^2}{2}  \int_0^{T} \!dt \: 
\sum_{\scriptstyle x,y\in \Lambda_N
\atop \scriptstyle |x-y|=1}  \eta_x(N^2t) \left[ 1- \eta_y(N^2t) \right]
 \left[ e^{H(t,y/N)-H(t,x/N) } - 1 \right]
\nonumber\\
&&\;\;\;\;\; - \frac{N^2}{2}  \int_0^{T} \!dt \: 
\sum_{\scriptstyle x\in \Lambda_N, y\not\in\Lambda_N
\atop \scriptstyle |x-y|=1}  
\eta_x(N^2t) \left[ e^{H(t,y/N)-H(t,x/N)} - 1 \right]
\nonumber\\
&&\;\;\;\;\; -\frac{N^2}{2}  \int_0^{T} \!dt \: 
\sum_{\scriptstyle x\in \Lambda_N, y\not\in\Lambda_N
\atop \scriptstyle |x-y|=1}  
\psi(y/N) \left[1-\eta_x(N^2t)\right] 
\left[ e^{H(t,x/N)-H(t,y/N)} - 1 \right]
\end{eqnarray}
we refer again to Appendix \ref{appa} for the derivation of the above
formula. 

By Taylor expansion, summation by parts, the replacements 
$\eta_x(N^2t)[1- \eta_y(N^2t)] \approx \rho_N(t,x/N)[1-\rho_N(t,x/N)]$
in the bulk and $\eta_x(N^2t) \approx \psi(t,x/N) / [1+\psi(t,x/N)]$
at the boundary (which can be heuristically justified by the same
argument given for the zero range process) we get 
\begin{eqnarray}
&&\mathcal{J}_{[0,T]}^N (\eta (\cdot),H)
\approx N^d \Big\{ \langle \rho_N(T), H(T) \rangle -
\langle \rho_N(0), H(0) \rangle 
\nonumber\\
&&\;\;\;\;\;
\left.
- \int_0^T\!dt \: 
\left\langle \rho_N(t), \partial_t H(t) +\frac 12 \Delta H(t)\right\rangle
- \frac 12 \int_0^{T} \!dt \: 
\left\langle \rho_N(t) [1-\rho_N(t)]\,, \left|\nabla H(t)\right|^2\right\rangle
\right.
\nonumber\\
&&\;\;\;\;\;
\left.
+ \frac 12 \int_0^{T} \!dt \: \int_{\partial\Lambda} \!du \: 
\frac{\psi(u)}{1+\psi(u)} \partial_{\hat n} H(t,u)
\right\}
\end{eqnarray}
Recalling that the hydrodynamic equation for the perturbed dynamics is
given by \reff{HsepP}, after an integration by parts, we finally get
the action functional $J$
\begin{equation}\label{Jsep}
J_{[0,T]} (\hat\rho) = \frac 12 \int_0^{T} \!dt \: 
\left\langle \hat\rho(t) [ 1-\hat\rho(t)]  \,, 
\left| \nabla H (t) \right|^2 \right\rangle
\end{equation}
which is of the form \reff{J2} with $D_{i,j}(\hat\rho)= \delta_{i,j}$
and $\chi_{i,j}(\hat\rho)= \hat\rho [ 1-\hat\rho]
\delta_{i,j}$. As for the zero range, the functional
$J_{[0,T]}(\hat\rho)$ is defined to be infinite if $\hat\rho$ does not
satisfy the boundary condition in \reff{HsepP}.

\ssubsection{Reduction of Hamilton--Jacobi to a non linear
differential equation ($d=1$)} 
\par\noindent 
We consider here the boundary driven simple exclusion process in one
space dimension. In a very interesting paper, by using a matrix representation
of the microscopic invariant measure and combinatorial techniques,
Derrida, Lebowitz and Speer \cite{DLS} have recently shown that the
action functional $S$ (which we called the macroscopic entropy) can be
expressed in terms of the solution of a non--linear ordinary
differential equation. We show next how this result can be recovered
in our approach by following the dynamical/variational route explained
in Section \ref{s:gt}. Namely, we consider the variational problem
\reff{l1} for the one--dimensional simple exclusion process and show
that the associated Hamilton--Jacobi equation which, taking into
account \reff{Jsep} and \reff{HsepP}, is the functional derivative
equation
\begin{equation}\label{HJsep}
\left\langle \nabla \frac{\delta S}{\delta\rho}, 
\rho (1-\rho)\nabla \frac{\delta S}{\delta\rho} \right\rangle
+\left\langle \frac{\delta S}{\delta\rho}, \Delta\rho\right\rangle = 0
\end{equation}
can be reduced to the non--linear ordinary differential equation first
obtained in \cite{DLS}.

For notation simplicity, we assume that $\Lambda= (-1,1)$, so that
$\partial\Lambda=\pm 1$. We shall also assume the macroscopic density
profile $\rho=\rho(u)$ satisfies the boundary conditions in equation
\reff{HE}. 

We look for a solution of the Hamilton--Jacobi equation \reff{HJsep}
by performing the change of variable
\begin{equation}\label{guess}
\frac{\delta S}{\delta\rho (u)} = 
\log \frac{\rho(u)}{1-\rho(u)} - \phi(u;\rho)
\end{equation}
for some functional $\phi(u; \rho)$ to be determined satisfying the
boundary conditions $\phi(\pm 1 )= \log \rho(\pm 1 )/[1-\rho(\pm 1)] 
=\log \psi(\pm 1)$.

Inserting \reff{guess} into \reff{HJsep}, we get (note that 
$\rho - e^\phi/{(1+e^\phi)}$ vanishes at the boundary)
\begin{eqnarray}\label{shjsep}
 0 &=& - \left\langle 
\nabla \left( \log \frac{\rho}{1-\rho} - \phi \right), \rho (1-\rho)
\nabla \phi \right\rangle
\nonumber\\
&=&- \left\langle \nabla\rho, \nabla\phi\right\rangle 
+ \left\langle \rho (1-\rho), (\nabla\phi)^2 \right\rangle
\nonumber\\
&=&-  \left\langle 
\nabla \left( \rho - \frac{e^\phi}{1+e^\phi} \right), \nabla \phi
\right\rangle
-\left\langle \left( \rho-\frac{e^\phi}{1+e^\phi}  \right)
\left(\rho-\frac{1}{1+e^\phi} \right), (\nabla\phi)^2 
\right\rangle
\nonumber\\
&=& \left\langle\left( \rho - \frac{e^\phi}{1+e^\phi} \right) , 
\left( \Delta \phi + \frac{ (\nabla\phi)^2}{1+e^\phi} 
- \rho (\nabla\phi)^2 \right)\right\rangle
\end{eqnarray}
We obtain a solution of the Hamilton--Jacobi if we solve the following
ordinary differential equation which relates the functional
$\phi(u)=\phi(u; \rho)$ to $\rho$
\begin{equation}\label{dphi}
\left\{
\begin{array}{l}{\displaystyle 
\frac {\Delta \phi(u)}{[\nabla\phi(u)]^2} + 
\frac{1}{1+e^{\phi(u)}} = \rho (u)  \quad u\in (-1,1) }
\\
\\
\phi(\pm 1) = \log \psi(\pm 1)
\end{array}
\right.
\end{equation}

A computation shows that the derivative of the functional 
\begin{equation}\label{Ssep}
S(\rho) = \int_{-1}^1\!du \left\{
\rho\log \rho + (1-\rho)\log(1-\rho) + (1-\rho) \phi 
- \log \left(1+e^{\phi}\right) + \log
\frac{\nabla\phi}{\nabla\bar\rho} 
\right\}
\end{equation}
is given by \reff{guess} when $\phi(u;\rho)$ solves \reff{dphi}.
According to the discussion in Section \ref{s:2.6}, we will be able to
conclude that \reff{Ssep} is indeed the macroscopic entropy as soon as
we show that it meets the criterion \reff{bf06}. This will be done in
the next Subsection.  By the change of variable $\phi = \log [ F /
(1-F)]$ equation \reff{dphi} becomes the one obtained in \cite{DLS}.

One may be tempted to repeat the same computation in arbitrary
dimension; one would obtain a partial differential equation analogous
to \reff{dphi}. However, in more than one dimension it does not exist,
in general, a functional $S$ whose derivative is given by \reff{guess}
with $\phi$ and $\rho$ related by such partial differential equation.

The equation \reff{dphi}, considered as a relationship expressing
$\rho$ in terms of $\phi$, can be interpreted in the following way. Let
\begin{equation}\label{S0sep}
S_0(\rho) = \int_{-1}^1\!du 
\left\{ \rho\log \rho + (1-\rho)\log(1-\rho) \right\}
\end{equation}
be the equilibrium entropy. 
Since $\delta S_0/\delta \rho = \log [\rho/(1-\rho)]$ we have
$$
\phi (u;\rho) = \frac {\delta S_0}{\delta \rho} 
- \frac {\delta S}{\delta \rho}
$$
If $\mathcal{G} (\phi)$ is the Legendre transform of $S_0-S$, we find
that $\delta \mathcal{G} /\delta \phi=\rho$ which is the relationship
\reff{dphi}.

We note that the remark after \reff{eqcomp} for the zero range
process also applies to the present context. In particular the
macroscopic computation \reff{shjsep} depends only on the action
functional $J$ and is therefore stable with respect to microscopic
perturbations which do not affect the dynamical large deviations.

\ssubsection{Adjoint hydrodynamics  ($d=1$)}
\par\noindent
By using the fluctuation dissipation relationship \reff{gfd} and the
expression \reff{guess} for $\delta S/ \delta\rho$, we find that the
adjoint hydrodynamics is given by the equation non local in space
\begin{equation}\label{adjHE}
\left\{
\begin{array}{l} 
{\displaystyle
\partial_t \rho^* (t,u) = \frac 12 \Delta \rho^* (t,u)  
- \nabla \left\{ \rho^*(t,u) [ 1- \rho^*(t,u)]  
\nabla \phi (u; \rho^*(t))
\right\} \quad u \in (-1,1)
}
\\
{\displaystyle \rho^*(t,\pm 1)   = \frac{\psi(\pm 1)}{1+\psi(\pm 1)} 
} 
\\
{\displaystyle \rho^*(0,u)   = \gamma(u) } 
\end{array}
\right.
\end{equation}
where $\phi (u; \rho^*(t))$ is to be obtained from $\rho^*(t)$ by
solving \reff{dphi}.  Since $\phi (u; \bar\rho) = \log [
\bar\rho/(1-\bar\rho)]$, we see that $\bar\rho$ is also a stationary
solution of \reff{adjHE}.

We now show how \reff{adjHE} can be related to the heat equation.  Let
$\rho^*(t,u)$ be the solution of \reff{adjHE} and introduce $F=F(t,u)$
as
\begin{equation}
\label{Fphi}
F(t,u)=  \frac { e^{\phi(u;\rho^*(t) )}}{1+e^{\phi(u;\rho^*(t) )}} 
\end{equation}
it is not too difficult to check (see Appendix \ref{s:ahs}) that
$F(t,u)$ satisfies the heat equation
\begin{equation}\label{adjHEF}
\left\{
\begin{array}{l} 
{\displaystyle
\partial_t F (t,u) = \frac 12 \Delta F (t,u)  
 \quad u \in (-1,1)
}
\\
{\displaystyle F(t,\pm 1)   = \frac{\psi(\pm 1)}{1+\psi(\pm 1)} 
} 
\\
{\displaystyle F(0,u)   =  
\frac { e^{\phi(u;\gamma )}}{1+e^{\phi(u;\gamma )}}}
\end{array}
\right.
\end{equation}
Conversely, given $F=F(t,u)$ which solves \reff{adjHEF}, by setting 
\begin{equation}\label{r*vF}
\rho^*(t,u) =  F(t,u) 
+  F(t,u) [1- F(t,u)]  \frac{ \Delta F(t,u)}{ [\nabla F(t,u) ]^2}
\end{equation}
a computation (see again Appendix \ref{s:ahs}) shows that $\rho^*$
solves \reff{adjHE}.

We have thus shown how a solution of the (non local, non linear)
equation \reff{adjHE} can be obtained from the linear heat equation by
performing the non local transformation \reff{Fphi} on the initial
datum.  In particular, since the solution $F(t,u)$ of \reff{adjHEF}
converges as $t\to\infty$ to $\bar\rho$, we see that the functional
$S(\rho)$ given in \reff{Ssep} satisfies the criterion \reff{bf06} so
that it is indeed the macroscopic entropy.

\ssubsection{Non perturbative lower bound on the macroscopic entropy
($d\ge 1$)}
\par\noindent
We discuss here a non perturbative bound for the macroscopic entropy
in the boundary driven simple exclusion process in arbitrary space
dimension $d$.  Let $S_0(\rho)$ be the equilibrium entropy as defined
in \reff{S0sep}, we shall obtain the following lower bound on
$S(\rho)$
\begin{eqnarray}\label{lbS}
S(\rho) &\ge &  S_{0}(\rho) - S_0(\bar\rho)  
- \left\langle \rho -\bar\rho ,\frac{\delta S_0}{\delta \rho}
(\bar\rho)\right\rangle
\nonumber \\
&=& \int_{\Lambda}\! du \: \Big\{ 
\rho \log \frac{\rho}{\bar\rho} +
(1-\rho) \log \frac{1-\rho}{1-\bar\rho}
\Big\} = {\tilde S} (\rho)
\end{eqnarray}
with a strict inequality for $\rho\neq\bar\rho$.
In the one dimensional case the bound \reff{lbS} has been independently
obtained in \cite{DLS}.

Recalling that the dynamical action functional $J$ of this model is
given by \reff{Jsep}, a somewhat lengthy but straightforward
computation gives
\begin{eqnarray}
\label{ricsquare}
J_{[-T,0]} (\hat\rho(\cdot)) &=&
\frac 12 \int_{-T}^0 \!dt \: 
\left\langle \nabla^{-1} \left( \partial_t \hat\rho  - 
\frac 12 \Delta \hat\rho \right), 
\frac{1}{\hat\rho [1-\hat\rho] } 
\nabla^{-1} \left( \partial_t  \hat\rho  - \frac 12 \Delta \hat\rho  
\right) \right\rangle
\nonumber\\
&=& {\tilde S} (\hat\rho(0)) -  {\tilde S} (\hat\rho({-T}))  
\nonumber\\
&& + \, \frac 12  \int_{-T}^0 \!dt \:
\left\langle \nabla^{-1} \left( \partial_t\hat\rho + \frac 12 
\Delta \hat\rho - \nabla \cdot \left( \hat\rho [ 1-\hat\rho] 
\nabla\log \frac{\bar\rho}{1-\bar\rho} \right)  \right) 
, 
\right.
\nonumber\\
&&\phantom{+}\, \left. 
\frac{1}{\hat\rho [1-\hat\rho] } 
\nabla^{-1} \left( \partial_t\hat\rho(t) + \frac 12 
\Delta \hat\rho - \nabla \cdot \left( \hat\rho [ 1-\hat\rho] 
\nabla\log \frac{\bar\rho}{1-\bar\rho} \right)  \right) 
\right\rangle
\nonumber\\
&& 
+\, \frac 12   \int_{-T}^0\!dt \int_\Lambda  \!du 
\: \frac{ \left| \nabla \bar\rho(u) \right|^2 }
     { [\bar\rho(u)(1-\bar\rho(u) )]^2}
\left(\hat\rho(t,u) -\bar\rho(u) \right)^2
\end{eqnarray}
The last two terms on the right hand side 
of (\ref{ricsquare}) are positive. Therefore, if $\hat\rho (t)$ is
trajectory connecting $\bar\rho$ to $\rho$, we have 
$$
S(\rho) = \inf_{\hat\rho} J_{[-\infty,0]} (\hat\rho)
\ge {\tilde S} (\rho) - {\tilde S} (\bar\rho) = {\tilde S}(\rho)
$$
Moreover, since the last term on the right hand side of
\reff{ricsquare} is strictly positive as soon as $\hat\rho \neq
\bar\rho$, we have the strict inequality in \reff{lbS} for
$\rho\neq\bar\rho$.

\ssubsection{Perturbative solution of the Hamilton--Jacobi equation
($d\ge 1$)} 
\par\noindent
We show here how the Hamilton--Jacobi equation (\ref{HJ}) can be used
to get a perturbative expansion for the entropy $S$ around the
stationary profile $\bar\rho$. We discuss only the expansion up to
the second order but it will be clear how to generate an iterative
approximation scheme. 

{}From a computational point of view it is convenient to expand the
pressure $G(h)$ defined in \reff{pressure}. Since $\rho (u) = \delta G
(h) / \delta h (u)$ and $h(u) =\delta S (\rho) / \delta \rho (u)$, the
dual Hamilton--Jacobi equation \reff{dualHJ} in this model is
\begin{equation}
\label{HJP}
\left\langle \nabla h, 
\left[ \frac{ \delta G}{\delta h} 
\left(1 - \frac{ \delta G}{\delta h}  \right)\right]
\nabla h \right\rangle
=\left\langle \nabla h , \nabla  \frac{ \delta G}{\delta h} \right\rangle
\end{equation}
where $h(u)$ satisfies the boundary conditions $h(u)=0$ for
$u\in\partial\Lambda$.  

Recall that $\bar\rho (u)$ is the stationary solution of the hydrodynamic
equation (\ref{HE}) and introduce
\begin{equation}
\label{Peq}
{\tilde G} (h) = \int_\Lambda \! du \: 
\log \left[ 1+ \bar\rho(u) \left( e^{h(u)} -1 \right)\right]
\end{equation}
Note that ${\tilde G} (h)$ is the Legendre transform of
${\tilde S} (\rho)$ defined in \reff{lbS}.
We look for a solution of \reff{HJP} in the form 
\begin{equation}\label{G=}
G(h) = {\tilde G} (h) + 
\langle g, h\rangle +\frac 12 \langle h, B h \rangle +
O(h^3) 
\end{equation}
for some function $g=g(u)$ and some linear operator $B$.

Note that $S(\rho)$ has a minimum for $\rho=\bar\rho$ and 
$$
S(\rho) = \frac 12 \left\langle \rho - \bar\rho, C^{-1} 
\left( \rho - \bar\rho \right) \right\rangle + 
O\left( \left( \rho - \bar\rho \right)^3 \right) 
$$
where $C$ is the covariance of the density fluctuations with respect
to the invariant measure. Therefore
\begin{equation}\label{G2}
G(h) = \langle \bar\rho, h \rangle 
+\frac 12 \left\langle h, C h \right\rangle + O(h^3)
\end{equation}
hence the second derivative of $G$ at $h=0$ is the covariance of the
density fluctuations. By comparing \reff{G=} to \reff{G2} we get
\begin{equation}
C =\frac {\delta^2 {\tilde G} }{\delta h^2}\Big|_{h=0} + B 
= \bar\rho (1-\bar\rho) \, \id + B 
\end{equation}
Since ${\tilde G}$ is the pressure for the equilibrium system the
operator $B$ represents the contribution to the covariance due to the
non equilibrium boundary conditions.  For the boundary driven simple
exclusion process the covariance of the fluctuation has been derived
in \cite{DFIP,S} where it is shown that it is non local in space. Therefore
the perturbative expansion of the Hamilton--Jacobi equation will give
a different derivation of the covariance.

We have 
\begin{eqnarray*}
&& \frac{ \delta {\tilde G} }{\delta h} 
\left(1 - \frac{ \delta {\tilde G} }{\delta h}  \right)
= \frac {\bar\rho (1-\bar\rho) e^h}{\left[ 1 + \bar\rho(e^h -1)  \right]^2}
\\
&&\nabla \frac{\delta  {\tilde G}  }{\delta h}  =
\frac {\bar\rho (1-\bar\rho) e^h}{\left[ 1 + \bar\rho(e^h -1)
\right]^2} \nabla h 
+ \frac {e^h }{\left[ 1 + \bar\rho(e^h -1)  \right]^2}\nabla\bar\rho
\end{eqnarray*}
so that by plugging \reff{G=} into \reff{HJP} and expanding up to
second order in $h$ we get
$$
\left\langle \nabla h , 
\left[ 1+ (1-2\bar\rho) h \right]\nabla\bar\rho  + \nabla g + \nabla B
h \right\rangle = 0
$$
Recalling that $h$ vanishes at the boundary, we thus get $g=0$ and 
$$
\left\langle \nabla \left( \frac {h^2}{2} \right),
(1-2\bar\rho) \nabla\bar\rho 
\right\rangle = \left\langle h, \Delta B h \right\rangle
$$
which, after an integration by parts, becomes 
(recall that $\Delta\bar\rho =0$)
$$
\left\langle h, \Delta B h \right\rangle = 
\left\langle h, \left|\nabla\bar\rho\right|^2    h \right\rangle
$$
The operator $B$ therefore satisfies
$$
\frac 12 \left[ \Delta B + B \Delta \right] 
= |\nabla \bar\rho|^2 
$$

In particular, if $\nabla\bar\rho$ is constant, $B$ has the kernel
\begin{equation}
B(u,v) =  |\nabla\bar\rho|^2 \Delta^{-1}(u,v)  
\end{equation}
where $\Delta^{-1}(u,v)$ is the Green function of the Laplacian (with
Dirichlet boundary conditions). The fact that $B$ is a negative
operator can also be obtained as a consequence of the bound $S(\rho)
\ge  {\tilde S} (\rho)$.

\smallskip
By analogous computation one can obtain also the higher order terms in
the expansion of the pressure which are the higher order cumulants.
In the one dimensional case $\nabla\bar\rho$ is constant and we state
below the result of the expansion up to the third order.
\begin{eqnarray*}
&&\!\!\! G(h)={\tilde G}(h) + \frac 12 \left|\nabla \bar\rho \right|^2
\left\langle h , \Delta^{-1}h \right\rangle 
\\
&&+ \frac 13 (\nabla \bar\rho)^2 
\left[ \left\langle h^2 
\left( 1 - \frac 23 \hat \rho \right), \Delta^{-1}h \right\rangle 
- \left\langle (\nabla h)^2 
\left( 1 - 2 \hat \rho \right), \Delta^{-2}h \right\rangle 
\right]
+  O (h^4)
\end{eqnarray*}

\appendice{Derivation of \protect\reff{strJ} 
and \protect\reff{strJsep}} 
\label{appa}
\par\noindent
Let $\Omega$ be a finite set and consider a continuous time jump Markov
process $X_t$ on the state space $\Omega$ with generator given by
\begin{equation}
\label{genas}
L f (\omega) = \sum_{\omega'\in\Omega} \lambda(\omega) 
p(\omega,\omega') \left[ f(\omega') - f (\omega) \right]
\end{equation}
where the rate $\lambda$ is a positive function on $\Omega$ and
$p(\omega,\omega')$ is a transition probability. 
We can construct a realization of $X_t$ as follows. 
Fix an initial condition $X_0=\omega_0$.  The process waits an
exponential time $\tau_{1}$ with rate $\lambda(\omega_0)$ and then  
jumps to $\omega_1$ with probability $p(\omega_0,\omega_1)$; the law
of $\tau_{1}$ is 
\begin{equation}
{ P}(\tau_{1}< t)= \int_0^t\!ds\: \lambda(\omega_0) \, 
e^{- \lambda(\omega_0) s}
\label{unk}
\end{equation} 
The process waits now an exponential time $\tau_{2}$, independent of
$\tau_1$, with rate $\lambda(\omega_1)$ and then jumps to $\omega_2$
with probability $p(\omega_1,\omega_2)$, and so on.  Consider the
piecewise constant trajectory $X_s$, $s\in [0,T]$ with $n$ jumps given
by
\begin{equation}
X_s = \left\{
\begin{array}{ll}
\omega_0 &      0\le  s < t_1 \\
\omega_1 &  t_1 \le s< t_1+ t_2 \\
\cdots & \cdots \\
\omega_{n-1} &  t_1 +t_2 +\cdots + t_{n-1} \le s <  t_1 +t_2 +\cdots +
t_{n} \\
\omega_n &  t_1 +t_2 +\cdots + t_{n} \le s \le T
\end{array}
\right.
\end{equation}
Its probability density is given by
$$
d{ P}_{\omega_0}(X_s,\, s\in [0,T] )= 
\prod_{i=1}^n  \Big( \lambda(\omega_{i-1})  
\, p \left(\omega_{i-1},\omega_{i}\right) 
\, e^{ - \lambda (\omega_{i-1}) t_i } \, dt_i
\Big) \cdot e^{-\lambda(\omega_n) [T-\sigma_n] }
$$
where $\sigma_n= t_1+\cdots +t_n$

If $\lambda$ and $p$ depend explicitly on time we can construct a
realization $X_t$ by the same procedure; in such a case the law of
$\tau_1$ is
\begin{equation}
{ P}(\tau_{1}< t)= \int_0^t\!ds\: \lambda(\omega_0,s) \, 
e^{- \int_0^s \!ds' \, \lambda(\omega_0, s')}
\end{equation}
and analogous distributions for $\tau_i$. We thus get
\begin{equation}\label{pdt}
d{ P}_{\omega_0}(X_s,\, s\in [0,T] )= 
\prod_{i=1}^n  \left( \lambda(X_{\sigma_{i-1}},\sigma_i)  
\, p \left(X_{\sigma_{i-1}}, X_{\sigma_i}; \sigma_i \right) \, dt_i 
\right) \cdot
e^{ - \int_{0}^{T} \! ds \, \lambda (X_s ,s) }
\end{equation}
where $\sigma_k=t_1+\cdots +t_k$ (resp. $\sigma_0=0$) are the jump
times of $X_s$.

Let us now consider two processes $X_t$ (resp. $X'_t$) of this type
with rates $\lambda(\omega,t)$ (resp. $\lambda'(\omega,t)$) and
transition probability $p(\omega,\omega';t)$
(resp. $p'(\omega,\omega';t)$).  We can write the formula \reff{pdt}
also for the process $X'$ and, by taking the ratio of those
expressions, we get the so called Radon--Nikodym derivative
\begin{equation}\label{RNd}
\begin{array}{l}
{\displaystyle
\frac{d{ P}_{\omega_0}'}{d{ P}_{\omega_0}}
(X_s, \, s\in [0,T] )=  
\exp\left\{ {\mathcal J}_{[0,T]} (X) \right\} 
}
\\ 
{\displaystyle \;\; = 
\exp\left\{ \sum_{i=1}^n \log 
\frac{\lambda'(X_{\sigma_{i-1}},\sigma_i) 
p'\left(X_{\sigma_{i-1}}, X_{\sigma_i}; \sigma_i \right) }
{\lambda(X_{\sigma_{i-1}},\sigma_i) 
p\left(X_{\sigma_{i-1}}, X_{\sigma_i}; \sigma_i \right) }
- \int_0^T\! ds \,
\big[ \lambda'(X_s,s) - \lambda(X_{s},s) \big]
\right\}
}
\end{array}
\end{equation}

We now consider the special case in which $X$ has generator given by
\reff{genas} and $X'=X^F$ has a time dependent generator
\begin{equation}
\label{genasF}
L^F_t f (\omega) = \sum_{\omega'\in\Omega} \lambda(\omega) 
p(\omega,\omega') e^{F(\omega',t)- F(\omega,t)}
\left[ f(\omega') - f (\omega) \right]
\end{equation}
which is of the same form with 
\begin{eqnarray*}
\lambda'(\omega,t) &=& \sum_{\omega'\in\Omega} 
\lambda(\omega) p(\omega,\omega') e^{F(\omega',t)- F(\omega,t)}
\\
p'(\omega,\omega';t) &=& \frac{1}{\lambda'(\omega,t)} 
\lambda(\omega) p(\omega,\omega') e^{F(\omega',t)- F(\omega,t)}
\end{eqnarray*} From \reff{RNd} we get that the Radon--Nikodym
derivative is given by  
$$
\frac{dP_{\omega_0}^F}{dP_{\omega_0}} 
(X_s, \, s\in [0,T] ) = \exp\{ {\mathcal J}_{[0,T]} (X,F)\}
$$
with
\begin{equation}\label{jjj}
\begin{array}{lll}
{\displaystyle \!\!\!
{\mathcal J}_{[0,T]} (X, F)
}
&=& {\displaystyle \sum_{i=1}^n \left[ F(X_{\sigma_i},\sigma_i) 
-  F(X_{\sigma_{i-1}},\sigma_i) \right]
}\\
&&
{\displaystyle - \int_0^T\! ds \,
\lambda(X_s) \sum_{\omega'\in\Omega} p(X_s,\omega') 
\left[ e^{F(\omega',s)- F(X_s,s)} - 1 \right]
}
\\
&=&
{\displaystyle 
\sum_{i=1}^n \left[ F(X_{\sigma_i},\sigma_i) -
  F(X_{\sigma_{i-1}},\sigma_{i-1}) 
- \int_{\sigma_{i-1}}^{\sigma_i} \!ds\: \partial_s  
F(X_{\sigma_{i-1}}, s ) \right] 
}\\
&&
{\displaystyle - \int_0^T\! ds \, e^{-F(X_s,s)} L e^{F(X_s,s)}
}\\
&=&
{\displaystyle 
F(X_T,T) - F(X_0,0) - \int_0^T \!ds 
\left[  \partial_s  F(X_s, s ) + e^{-F(X_s,s)} L e^{F(X_s,s)} \right]
}
\end{array}
\end{equation}

Formulae \reff{strJ} and \reff{strJsep} are special cases of
\reff{jjj} obtained by choosing
$$
F(\eta,\tau) = \sum_{x\in\Lambda_N} H(\tau/N^2, x/N) \eta_x
$$

\appendice{Adjoint hydrodynamics for one dimensional simple
exclusion}\label{s:ahs} 
\par\noindent
Let $\rho^*(t)$ be a solution of the adjoint hydrodynamics for the 
one dimensional simple exclusion process \reff{adjHE}. 
By the remarks in Subsection \ref{s:hi}, $\big(\rho(t), H(t) \big)$
with 
\begin{eqnarray}\label{rH}
\rho(t)&=& \rho^*(-t) 
\nonumber \\
H(t) & =&  \frac{\delta S}{\delta\rho} \big( \rho^*(-t) \big)
= \log \frac{\rho^*(-t)}{1- \rho^*(-t)} -\phi(\rho^*(-t))
\end{eqnarray}
is a solution of the Hamilton equations \reff{hameq} which for this
model read
\begin{eqnarray}\label{Hesep}
\partial_t \rho &=& \frac 12 \Delta \rho 
		- \nabla \Big( \rho(1-\rho) \nabla H \Big) 
\nonumber\\
\partial_t H &=& - \frac 12 (1-2\rho) \big( \nabla H\big)^2  
		-\frac 12 \Delta H
\end{eqnarray}
By plugging \reff{rH} into \reff{Hesep} and performing the change of
variable \reff{Fphi}, a straightforward computation yields
\begin{eqnarray}\label{Hesused}
\partial_t \rho^* &=& \frac 12 \Delta \rho^* 
- \nabla \left( \frac {\rho^*(1-\rho^*)}{F(1-F)} \nabla F \right) 
\nonumber\\
\partial_t F &=& - \frac 12 \Delta F
+ (\rho^* - F) \frac {\big( \nabla F\big)^2}{F(1-F)}  
\end{eqnarray}
By writing the equation \reff{dphi} in terms of $F= e^{\phi}/
(1+e^{\phi})$ and $\rho$ replaced by $\rho^*$ we get
\begin{equation}\label{derridaeq}
\left\{
\begin{array}{l}{\displaystyle 
\big( \rho^* - F \big) = F(1-F) \frac {\Delta F}{ \big(\nabla F\big)^2 } 
\quad \textrm{ for any  } (t,u) \in [0,\infty) \times (-1,1) }
\\
\\
F(t,\pm 1) =\rho^*(t,\pm 1) = \frac {\psi(\pm 1)}{1+\psi(\pm 1)}
\end{array}
\right.
\end{equation}
which inserted in \reff{Hesused} concludes the proof that $F(t,u)$
as defined in \reff{Fphi} satisfies the heat equation.

The converse statement, namely that if we define $\rho^*=\rho^*(t,u)$
as in \reff{r*vF} (with $F=F(t,u)$ the solution of \reff{adjHEF}) then it 
satisfies the non local equation \reff{adjHE}, can be checked without
invoking the Hamiltonian formalism. 
Indeed, from \reff{r*vF} we get that
\begin{equation}\label{r1-r}
\frac{ \rho^*(1-\rho^*)}{ F(1-F)} = 
1 + (1-2F) \frac{\Delta F}{ \big(\nabla F\big)^2} 
- F(1-F) \frac{\big(\Delta F\big)^2}{\big(\nabla F\big)^4}
\end{equation}
recalling \reff{adjHEF}, by a somehow tedious computation of the partial
derivatives which we omit, we get
\begin{equation}\label{fino4derivata}
\left( \partial_t - \frac 12 \Delta \right) 
\left[ F(1-F) \frac {\Delta F}{\big(\nabla F\big)^2} \right]
= - \nabla \Big( \frac{ \rho^*(1-\rho^*)}{ F(1-F)} \nabla F \Big)
\end{equation}
Therefore, recalling \reff{r*vF}, the function
$\rho^*(t)$ satisfies 
\begin{equation}\label{adjhconF}
\partial_t \rho^* = \frac 12 \Delta \rho^* 
- \nabla \left( \frac {\rho^*(1-\rho^*)}{F(1-F)} \nabla F \right)
\end{equation}
which is precisely \reff{adjHE} written in terms of the variable $F=F(\rho^*)$
instead of $\phi=\phi(\rho^*)$.

\bigskip\bigskip\noindent 
\normalsize\textbf{Acknowledgments}\par
\smallskip 
\par\noindent
{\small We are grateful to P. Butt\`a, A. De Masi, P. Ferrari,
G. Giacomin, E. Presutti, and M.E. Vares for very illuminating
discussions.  We also thank B. Derrida, J.L. Lebowitz, and E.R. Speer
for communicating to us their work before publication. 
D. G. was supported by FAPESP 98/11899-2 at IME-USP
(Brazil) and, in the final stage of the work, by the Wittgenstein
award (P. Markowich).  L. B., G. J.-L. and C. L. acknowledge the
support of Cofinanziamento MURST.  }

\end{document}